\documentclass[12pt]{article}

\usepackage{amssymb}

\makeatletter
 \@addtoreset{equation}{section}
 
\makeatother

\newcommand{\tm}{\tilde{m}}
\newcommand{\tn}{\tilde{n}}

\newcommand{\ta}{\tilde{a}}
\newcommand{\tb}{\tilde{b}}

\newcommand{\tmu}{\tilde{\mu}}
\newcommand{\tnu}{\tilde{\nu}}

\newcommand{\ha}{\hat{a}}
\newcommand{\hb}{\hat{b}}
\newcommand{\hc}{\hat{c}}
\newcommand{\hd}{\hat{d}}
\newcommand{\he}{\hat{e}}

\newcommand{\hap}{a_{\perp}}
\newcommand{\hbp}{b_{\perp}}
\newcommand{\hcp}{c_{\perp}}

\newcommand{\hm}{\hat{m}}
\newcommand{\hn}{\hat{n}}
\newcommand{\hp}{\hat{p}}

\newcommand{\hmp}{m_{\perp}}
\newcommand{\hnp}{n_{\perp}}
\newcommand{\hpp}{p_{\perp}}
\newcommand{\hqp}{q_{\perp}}

\newcommand{\dhalpha}{\dot{\hat{\alpha}}}

\newcommand{\bpu}{\dot{+}}
\newcommand{\bmi}{\dot{-}}
\newcommand{\br}{\bar{r}}

\newcommand{\hga}{\hat{\gamma}'}
\newcommand{\lamm}{\Theta_{-}^{(+)}}
\newcommand{\blamm}{\bar{\Theta}^{(+)}_{-}}
\newcommand{\psm}{\Theta^{(-)}_{-}}
\newcommand{\bpsm}{\bar{\Theta}^{(-)}_{-}}

\newcommand{\metric}{\mbox{\boldmath$h$}}

\begin{document}


\begin{titlepage}

\renewcommand{\thefootnote}{\fnsymbol{footnote}}

\begin{flushright}
MIFP-04-10 \\
hep-th/0405260 \\
May 2004
\end{flushright}

\bigskip

\begin{center}
{\large \bf Matrix Quantum Mechanics for Supermembrane
   on $AdS_{7} \times S^{4}$}
\end{center}

\bigskip

\bigskip

\begin{center}
{\large Koichi Murakami}\footnote{e-mail:
        kmurakami@physics.tamu.edu}
\end{center}

\bigskip

\begin{center}
{\it
George P.\ and Cynthia W.\ Mitchell Institute for Fundamental
Physics,\\
Texas A\&M University, College Station, TX 77843-4242,
USA}
\end{center}

\bigskip

\bigskip

\bigskip

\begin{abstract}
We explore the light-cone gauge formulation of a closed
supermembrane on $AdS_{7} \times S^{4}$. We obtain the action
of matrix quantum mechanics with large $N$ $U(N)$ gauge symmetry
for the light-cone supermembrane. We show that this action
reproduces leading order terms in $\alpha'$-expansion of
the non-abelian Born-Infeld action of $N$ D$0$-branes
propagating near the horizon of D$4$-branes. The matrix quantum
mechanics obtained in this paper, therefore, has an interpretation
as Matrix theory in the near-horizon of D$4$-branes.
\end{abstract}

\setcounter{footnote}{0}
\renewcommand{\thefootnote}{\arabic{footnote}}

\end{titlepage}

\section{Introduction}

Supermembranes \cite{BST} play  important roles in M-theory.
{}For example, they lead to the equations of motion of
the low-energy effective field theory of M-theory
(i.e.\ eleven-dimensional supergravity) \cite{BST}\cite{DHIS},
provide string states in type IIA string theory \cite{DHIS}
and in $E_{8} \times E_{8}$ heterotic string theory
\cite{Horava-Witten}, and
become D$2$-branes in type IIA string theory
\cite{Townsend}\cite{Schmidhuber}.
Besides these facts, in Matrix theory \cite{BFSS},
infinitely many D$0$-branes are conjectured to
capture all physical degrees of freedom of M-theory in the
infinite momentum frame and the resulting D$0$-brane action
 \cite{bound}\cite{Tseytlin}
coincides with that of a closed light-cone supermembrane
in the flat space \cite{dWHN}.
Supermembranes are therefore expected to play
a pivotal role in pursuit of the microscopic description of
M-theory.

Soon after Matrix theory was proposed, its curved-space
extension was investigated from the viewpoint of D$0$-brane
physics in
e.g.\ refs.\cite{Douglas}\cite{Douglas-Ooguri}\cite{Taylor}.
We may naturally expect that
as well as these analyses, the investigation of light-cone
supermembranes in curved spaces should provide another promising
approach to this problem,
taking into account the above-mentioned coincidence
in the flat space between
the action of D$0$-branes and that of a light-cone supermembrane.
The analyses along this line were initiated
by ref.\cite{dWPP-curve}.

In ref.\cite{dWPP-curve}, quite general results are obtained
regarding the actions of bosonic light-cone membranes
on curved backgrounds
and those of the corresponding matrix theories,
and then contributions of the fermionic coordinate $\Theta$
of supermembranes are included up to quadratic order.
The actions of supermembranes in curved spaces
are described in terms of the supervielbeins and the three-form
superfields in the eleven-dimensional $\mathcal{N}=1$
superspace~\cite{BST}.
Since the fermionic coordinate $\Theta$
in the superspace is an $SO(10,1)$ Majorana spinor,
the superfields are, in general,
$32$nd order polynomials of $\Theta$.
In generic situation, it becomes an intricate task
to determine the superfields in all order
of $\Theta$ \cite{dWPP-curve}.
This fact leads us to specific cases
in which the backgrounds possess large isometry,
because the restriction of the symmetry is expected to
facilitate the task.
In fact, for the maximally supersymmetric solutions
of eleven-dimensional supergravity:
Minkowski space, $AdS_{4} \times S^{7}$,
$AdS_{7} \times S^{4}$ and the pp-wave solution,
the full-order forms of the supervielbeins,
super spin-connections
and the three-form superfields are determined.
In the $AdS_{4} \times S^{7}$ and the $AdS_{7} \times S^{4}$
cases, the superfields are obtained
in refs.\cite{KRR}\cite{dWPPS}\cite{Claus}.
These superfields are reduced to those of
the pp-wave background
in the Penrose limit \cite{Penrose}\cite{Guven-P}.

Related to the flat case,
in refs.\cite{Russo-Tseytlin}\cite{Sekino-Yoneya}\cite{Huang},
a light-cone supermembrane is considered
in the eleven-dimensional Minkowski space
with non-trivial periodicity from which the ten-dimensional
Kaluza-Klein Melvin background is derived.
In refs.\cite{DSJR}\cite{Sugiyama-Yoshida},
the light-cone gauge formulation of a supermembrane
on the pp-wave background
is constructed and it is shown that the resulting system becomes
the matrix model proposed by Berenstein, Maldacena and Nastase
(BMN) \cite{BMN}.
Several properties of the BMN matrix model are studied also
by refs.\cite{N-S-Y}\cite{MSJR}\cite{Kimura-Yoshida}.

In this paper, we study the light-cone gauge formulation
of a closed supermembrane on $AdS_{7} \times S^{4}$.
Similarly to the flat and the pp-wave cases,
we obtain the matrix quantum mechanics for the supermembrane.
We show that the matrix action of this system
takes the same form as that of D$0$-branes propagating
near the horizon of D$4$-branes.
This result is consistent with the fact
that $AdS_{7} \times S^{4}$ is obtained as near-horizon geometry
\cite{Gibbons-Townsend}\cite{Maldacena}
of the M$5$-brane solution.

M-theory on $AdS_{7}\times S^{4}$ is proposed to
be dual to six-dimensional superconformal field theory
on the M$5$-brane world-volume
\cite{Witten95}\cite{Strominger}\cite{Witten5-brane}\cite{Seiberg}.
This is an example of the $AdS/CFT$ correspondence conjectured by
Maldacena \cite{Maldacena}.
To obtain insights into this $AdS_{7}/CFT_{6}$ correspondence,
(semi-)classical analyses of supermembranes on $AdS_{7}\times S^{4}$
are carried out in
refs.\cite{giant-graviton}\cite{Sezgin-Sundell}
\cite{Alishahiha}\cite{Bozhilov}.
Our result should give a new approach to the study
of this correspondence.

We make a comment on light-cone supermembranes on
$AdS_{4} \times S^{7}$.
In the $AdS_{4} \times S^{7}$ solution,
the three-form gauge potential $C_{\hm\hn\hp}$
does not satisfy the relation $C_{+-\hm}=0$.
Hence, the $X^{-}$-dependence is not straightforwardly
eliminated from the
light-cone hamiltonian of the supermembrane \cite{dWPP-curve}.
It might deserve more study \cite{dWPP-curve}.
In this paper, we will not  purse this case further.

It is worth mentioning that light-cone Green-Schwarz superstrings
on $AdS_{5} \times S^{5}$ are studied
\cite{Pesando}\cite{MT-LC}\cite{MTT}.
Our investigation is therefore an extension of these string analyses
into the supermembrane case.

This paper is organized as follows:
In Section \ref{sec:LCF}, we construct light-cone gauge
formulation of a supermembrane on $AdS_{7} \times S^{4}$.
The resulting action may be regarded as that of quantum mechanics
with area-preserving diffeomorphism gauge symmetry.
We employ the matrix regularization
\cite{Hoppe}\cite{dWHN}\cite{FFZ} for this action.
The system consequently becomes matrix quantum mechanics
with large $N$ $U(N)$ gauge symmetry.
In Section \ref{sec:D0PNHD4}, we show that the matrix action
obtained in Section \ref{sec:LCF} coincides with that of
large $N$ D$0$-branes propagating near the horizon of
D$4$-branes.
This implies that the light-cone supermembrane
on $AdS_{7} \times S^{4}$ has an interpretation in Matrix theory,
similarly to the flat case.
Section \ref{sec:summary} is devoted to conclusions and
discussions.
In Appendix \ref{sec:gamma}, we provide conventions of
the $SO(10,1)$ gamma matrices and spinors used in this paper.
In Appendices \ref{sec:derivation} and \ref{sec:note},
rearrangements and detailed forms of several equations
are presented.

\section{Light-Cone Gauge Formulation}
\label{sec:LCF}

\subsection{Supermembrane action on $AdS_{7} \times S^{4}$}

The action of a supermembrane on  curved space-time
is given in ref.\cite{BST} as
\begin{equation}
S= \int d^{3} \xi \, \mathcal{L}
=-T \int d^{3} \xi \left[
   \sqrt{-\det h_{ij}}
   + \frac{\epsilon^{ijk}}{3!} \Pi^{A}_{i} \Pi^{B}_{j} \Pi^{C}_{k}
   C_{ABC} \right]~,
 \label{eq:action1}
\end{equation}
where $T$ is the membrane tension,
which is described in terms of the eleven-dimensional
Planck length $l_{p}$ as
\begin{equation}
T=\frac{1}{(2\pi)^{2} l_{p}^{3}}~,
 \label{eq:m2tension}
\end{equation}
$\xi^{i}=(\tau,\sigma^{\acute{\imath}})$
$(\acute{\imath}=1,2)$ are the world-volume coordinates
on the membrane, and $\epsilon^{ijk}$  and  $h_{ij}$ 
are respectively the anti-symmetric tensor density and
 the induced metric
on the world-volume defined as
\begin{equation}
\epsilon^{\tau\sigma^{1}\sigma^{2}} = -1~;
\qquad
h_{ij} = \Pi_{i}^{\ha} \Pi_{j}^{\hb} \eta_{\ha \hb}~,
\quad
\Pi^{A}_{i} \equiv \partial_{i} Z^{M} E_{M}^{A}
\quad (A=\hat{a},\hat{\alpha};\ M=\hat{m},\dot{\hat{\alpha}})~.
\end{equation}
Here $Z^{M}=(X^{\hm}, \Theta^{\dhalpha})$ are the coordinates
on the eleven-dimensional curved superspace,
$E^{A}_{M}$ denotes the supervielbein
and $\eta_{\ha \hb}$ is the eleven-dimensional Minkowski metric
in the local Lorentz frame.
The fermionic coordinate $\Theta^{\hat{\alpha}}$
is a Majorana spinor of the $SO(10,1)$ local Lorentz group.
The conventions of $SO(10,1)$ spinors and the gamma matrices
used in this paper are presented in Appendix \ref{sec:gamma}.
The three-form
\begin{equation}
C^{(3)}=\frac{1}{3!}\Pi^{A} \wedge \Pi^{B} \wedge \Pi^{C} C_{ABC}~,
\quad
\Pi^{A} = dZ^{M} E_{M}^{A}~,
\end{equation}
is the potential for the four-form field-strength
\begin{equation}
F^{(4)}= dC^{(3)} = \frac{1}{4!} \Pi^{A}\wedge \Pi^{B}\wedge
 \Pi^{C} \wedge \Pi^{D}  F_{ABCD}~.
\end{equation}
In the superspace geometry of
eleven-dimensional supergravity,
there exists a closed four form
\begin{equation}
F^{(4)}=\frac{1}{4!}
 \left( \Pi^{\ha} \wedge \Pi^{\hb} \wedge \Pi^{\hc}
        \wedge \Pi^{\hd} F_{\ha \hb \hc \hd}
        +12 i \bar{\Pi} \wedge \Gamma_{\ha \hb} \Pi
             \wedge \Pi^{\ha} \wedge \Pi^{\hb} \right)~.
\label{eq:4-form}
\end{equation}

We consider the $AdS_{7} \times S^{4}$ solution of
eleven-dimensional supergravity.
The non-vanishing components of the Riemannian tensor and the
four-form field-strength of this solution are
\begin{eqnarray}
R_{abcd}&=&-\frac{1}{L^{2}} \left(
  \eta_{ac}\eta_{bd}-\eta_{ad}\eta_{bc}  \right)~, \nonumber\\
R_{a'b'c'd'}&=&\frac{4}{L^{2}}
   \left( \delta_{a'c'} \delta_{b'd'} - \delta_{a'd'} \delta_{b'c'}
   \right)~, \nonumber\\
F_{a'b'c'd'} &=& -\eta \frac{6}{L} \epsilon_{a'b'c'd'}~,
\quad \eta=\pm 1~,
\label{eq:ads7-sol}
\end{eqnarray}
where $a,b,c,d$ and $a', b', c', d'$ are local Lorentz indices
in the $AdS_{7}$ and the $S^{4}$ directions respectively,
and $L$ is the radius of $AdS_{7}$,
which is related to the radius $L_{S^{4}}$ of $S^{4}$
by $L_{S^{4}}=\frac{1}{2} L$.
Via a supercoset construction,
the supervielbein one-form
$\Pi^{A}=dZ^{M}E_{M}^{A}$
 and the superconnection one-form
$\Omega^{\ha \hb}=dZ^{M}{\Omega_{M}}^{\ha\hb}$
for this background are obtained \cite{KRR}\cite{dWPPS}
 in the all  order of the fermionic coordinate $\Theta$
as\footnote{The same result as eq.(\ref{eq:fullorder})
is obtained in ref.\cite{Claus} by using only eleven-dimensional
supergravity torsion and curvature constraints.}
\begin{eqnarray}
\Pi^{\hat{\alpha}} &=&
  \sum_{k=0}^{15} \frac{1}{(2k+1)!}
   {(\mathcal{M}^{2k})^{\hat{\alpha}}}_{\hat{\beta}}
   \, (\tilde{D} \Theta)^{\hat{\beta}}~, \nonumber\\
\Pi^{\ha} &=& e^{\ha}
  -2i \sum_{k=0}^{15} \frac{1}{(2k+2)!}
   \bar{\Theta} \Gamma^{\ha} \mathcal{M}^{2k} \tilde{D} \Theta~,
  \nonumber\\
\Omega^{\ha\hb} &=& \omega^{\ha\hb}
 +\frac{i}{72} \sum_{k=0}^{15} \frac{1}{(2k+2)!}
  \bar{\Theta} \left(\Gamma^{\ha\hb\hc_{1}\hc_{2}\hc_{3}\hc_{4}}
                     F_{\hc_{1}\hc_{2} \hc_{3} \hc_{4}}
               +24 \Gamma_{\hc\hd} F^{\ha\hb\hc\hd} \right)
  \mathcal{M}^{2k} \tilde{D} \Theta~.
\label{eq:fullorder}
\end{eqnarray}
Here $e^{\ha}=dX^{\hm}e^{\ha}_{\hm}$ and
$\omega^{\ha\hb}=dX^{\hm}{\omega_{\hm}}^{\ha\hb}$ are
the vielbein and the spin-connection one-forms
of $AdS_{7} \times S^{4}$ respectively,
and the one-form $\tilde{D} \Theta$
and the matrix $\mathcal{M}^{2}$ are defined as
\begin{eqnarray}
\tilde{D} \Theta &=&
 \left(d + \frac{1}{4} \omega^{\ha\hb} \Gamma_{\ha\hb}
       +e^{\ha}{T_{\ha}}^{\hb \hc \hd \he}
        F_{\hb \hc \hd \he} \right) \Theta~,
    \label{eq:d-theta}
        \\
{(\mathcal{M}^{2})^{\hat{\alpha}}}_{\hat{\beta}}
   &=& -2 i \left( {T_{\ha}}^{\hb \hc \hd \he} F_{\hb \hc \hd \he}
                   \Theta \right)^{\hat{\alpha}}
            \left(\bar{\Theta} \Gamma^{\ha} \right)_{\hat{\beta}}
     \nonumber\\
    &&
       {}+\frac{i}{288} \left(\Gamma_{\ha \hb}\Theta
                      \right)^{\hat{\alpha}}
        \left(\bar{\Theta} \Gamma^{\ha \hb \hc_{1} \hc_{2}\hc_{3}\hc_{4}}
         F_{\hc_{1} \hc_{2}\hc_{3}\hc_{4}}
         +24 \bar{\Theta} \Gamma_{\hc \hd} F^{\ha\hb\hc\hd}
          \right)_{\hat{\beta}}~,
    \label{eq:m-2}
\end{eqnarray}
where ${T_{\ha}}^{\hb \hc \hd \he}$ denotes
\begin{equation}
{T_{\ha}}^{\hb \hc \hd \he}
 = \frac{1}{288} \left( {\Gamma_{\ha}}^{\hb\hc\hd\he}
     -8 \delta_{\ha}^{[\hb}\Gamma^{\hc\hd\he]} \right)~.
\end{equation}
{}For the $AdS_{7}\times S^{4}$ solution,
the closed four-form (\ref{eq:4-form}) is integrated and
the three-form potential $C^{(3)}$ is obtained explicitly 
\cite{dWPPS} as follows:
\begin{equation}
C^{(3)}(Z) = \frac{1}{3!} e^{\ha}\wedge e^{\hb} \wedge e^{\hc}\,
                C_{\ha \hb \hc} (X) 
                + i \int^{1}_{0} dt\, \bar{\Theta}
              \Gamma_{\ha \hb} \Pi (X,t\Theta)
              \wedge \Pi^{\ha}(X,t\Theta) \wedge \Pi^{\hb}(X,t\Theta)~.
\end{equation}

\subsection{Light-cone gauge supermembrane}
  \label{sec:LCgauge}

In this subsection, we impose the light-cone gauge conditions
to fix the reparametrization invariance on the world-volume and
the fermionic gauge symmetry ($\kappa$-symmetry) \cite{BST}
of the action (\ref{eq:action1}).

We choose the $AdS_{7} \times S^{4}$ coordinates
in terms of which the metric
 takes the form
\begin{equation}
ds^{2}=
  G_{\hm \hn} dX^{\hm} dX^{\hn}
= \frac{2r}{L} \eta_{\tmu\tnu} dX^{\tmu}dX^{\tnu}
        +\left(\frac{L}{2r} \right)^{2} dr^{2}
        +\frac{L^{2}}{4}d\Omega_{4}^{2}~,
 \label{eq:metric1}
\end{equation}
where $(X^{\tmu},r)$ $(\tmu=0,\ldots,4,\natural)$
 are the horospherical
coordinates of $AdS_{7}$ and
$d\Omega_{4}^{2}$ is the metric of a unit $S^{4}$.
We denote the coordinates on $S^{4}$ by $X^{m'}$
$(m'=6,7,8,9)$.
The metric (\ref{eq:metric1}) can directly be obtained
as near-horizon geometry \cite{Gibbons-Townsend}\cite{Maldacena}
of the M$5$-brane solution \cite{Guven}
of eleven-dimensional supergravity,
\begin{equation}
ds^{2}_{M5}=\left(1+\frac{Ql_{p}^{3}\pi}{r^{3}}\right)^{-\frac{1}{3}}
             \eta_{\tmu\tnu} dX^{\tmu}dX^{\tnu}
             +\left(1+\frac{Ql_{p}^{3}\pi}{r^{3}}\right)^{\frac{2}{3}}
              \left(dr^{2}+r^{2} d\Omega_{4}^{2}\right)~,
   \label{eq:M5metric}
\end{equation}
where $Q$ denotes the number of the M$5$-branes,
$X^{\tmu}$ denote the directions parallel to the M$5$-branes
and $(r,X^{m'})$ denote the polar coordinates parametrizing
the directions perpendicular to the M$5$-branes.
The near-horizon limit taken in ref.\cite{Maldacena} is 
\begin{equation}
l_{p} \rightarrow 0
\qquad \mbox{with \ \ $\displaystyle \frac{r}{l_{p}^{3}}=$ finite}~.
\label{eq:nh-limit}
\end{equation}
One can  find that the radius of the $AdS_{7}$
is expressed \cite{Maldacena} as
\begin{equation}
L=2l_{p}(\pi Q)^{\frac{1}{3}}~.
\label{eq:adsradius}
\end{equation}

We introduce the light-cone coordinates
$X^{\hm}=(X^{+},X^{-},X^{\hmp})$ with
\begin{equation}
X^{\pm} = \frac{1}{\sqrt{2}} (X^{\natural} \pm X^{0})~,
\quad
X^{\hmp}=(X^{\tm},r,X^{m'})
\quad
  \tm=(1,\ldots, 4)~.
\label{eq:LC}
\end{equation}
We denote the local Lorentz indices corresponding to these coordinates
by $\ha=(\bpu, \bmi, \hap)$ with
$\hap=(\ta,\br,a')$ and
$\ta=(\bar{1},\ldots,\bar{4})$.
In terms of the coordinates (\ref{eq:LC}),
the metric (\ref{eq:metric1}) is recast into
\begin{equation}
ds^{2} = \frac{2r}{L}
 \left(2dX^{+}dX^{-}+ \delta_{\tm \tn}dX^{\tm} dX^{\tn}\right)
 + \left(\frac{L}{2r} \right)^{2} dr^{2}
 +\frac{L^{2}}{4} d\Omega^{2}_{4} ~.
 \label{eq:metricLC}
\end{equation}
We may choose the vielbein of $AdS_{7}$ in a diagonal form,
\begin{equation}
e^{\bpu}_{+} = e^{\bmi}_{-} = e^{\bar{1}}_{1}
 = e^{\bar{2}}_{2} =e^{\bar{3}}_{3}
 = e^{\bar{4}}_{4}=\sqrt{\frac{2r}{L}}~,
\quad e^{\br}_{r}= \frac{L}{2r}~.
\label{eq:vielbein-ads7}
\end{equation}
The non-vanishing elements of the spin-connection in the
$AdS_{7}$ sector
become
\begin{equation}
{\omega_{+}}^{\bpu\br}={\omega_{-}}^{\bmi \br}
={\omega_{1}}^{\bar{1}\br}
={\omega_{2}}^{\bar{2}\br}
={\omega_{3}}^{\bar{3}\br}
={\omega_{4}}^{\bar{4}\br}
=\sqrt{\frac{2r}{L^{3}}}~.
\label{eq:spinconnection-ads7}
\end{equation}
This yields
${\omega_{+}}^{\bpu \br} = \frac{1}{L} e^{\bpu}_{+}$,
${\omega_{-}}^{\bmi \br}= \frac{1}{L} e^{\bmi}_{-}$
and
${\omega_{\tm}}^{\ta \br}= \frac{1}{L} e^{\ta}_{\tm}$.

In order to fix gauge symmetries of the supermembrane action
(\ref{eq:action1}), we impose the
conditions\footnote{In this paper, we use the convention
in which the world-volume time coordinate $\tau$ has dimensions
of $(\mbox{length})^{1}$ in the space-time sense.
This convention is possible because of the invariance of
the action (\ref{eq:action1}) under the transformation
$\xi^{i} \mapsto l^{(i)} \xi^{i}$ for arbitrary constants $l^{(i)}$,
which is a part of the reparametrization invariance of the membrane
world-volume.}
\begin{equation}
X^{+}= \tau~, \quad \Gamma^{\bpu} \Theta =0~.
\label{eq:LCgauge1}
\end{equation}
Introducing the projection operator
$\mathcal{P}^{\mathrm{(LC)}}_{\pm}
  =\frac{1}{2} (1 \pm \Gamma^{\bmi\bpu} )$,
we separate the fermionic coordinate $\Theta$ into two
pieces:
\begin{equation}
\Theta= \Theta_{+}+\Theta_{-},
\quad
\Theta_{\pm} = \mathcal{P}^{(\mathrm{LC})}_{\pm} \Theta~.
\end{equation}
The gauge condition (\ref{eq:LCgauge1})
imposed on $\Theta$ is equivalent to
\begin{equation}
\Theta_{+} =0~.
\label{eq:LCfermi}
\end{equation}
We note that the `chirality' associated with the projection operator
$\mathcal{P}^{(\mathrm{LC})}_{\pm}$
flips under the Dirac conjugation:
\begin{equation}
\bar{\Theta}_{\pm}=\bar{\Theta}_{\pm} \mathcal{P}^{\mathrm{(LC)}}_{\mp}~.
\end{equation}

Applying the condition (\ref{eq:LCfermi}) to eq.(\ref{eq:m-2}),
we obtain
\begin{eqnarray}
{\mathcal{M}^{2}[\Theta_{-}]^{\hat{\alpha}}}_{\hat{\beta}}
&=& {}-2i \left( {T_{\hap}}^{b'c'd'e'}F_{b'c'd'e'} \Theta_{-}
    \right)^{\hat{\alpha}}
    \left(\bar{\Theta}_{-} \Gamma^{\hap} \right)_{\hat{\beta}}
  +\frac{i}{144} \Theta^{\hat{\alpha}}_{-}
    \left(\bar{\Theta}_{-} \Gamma^{a'b'c'd'}F_{a'b'c'd'}
    \right)_{\hat{\beta}}
   \nonumber\\
&&{}+\frac{i}{288} \left(\Gamma_{\hap\hbp} \Theta_{-}
      \right)^{\hat{\alpha}}
      \left( \bar{\Theta} \Gamma^{\hap\hbp c'_{1}c'_{2}c'_{3}c'_{4}}
              F_{c'_{1}\cdots c'_{4}}
      \right)_{\hat{\beta}}\nonumber\\
&&{}
   +\frac{i}{12} \left(\Gamma_{a'b'}\Theta_{-}\right)^{\hat{\alpha}}
       \left(\bar{\Theta}_{-} \Gamma_{c'd'} F^{a'b'c'd'}
       \right)_{\hat{\beta}}~.
    \label{eq:M2-LC}
\end{eqnarray}
Here we have used the relations
\begin{equation}
\Gamma^{\bpu}\Theta_{-}=\bar{\Theta}_{-}\Gamma^{\bpu}=0~.
\label{eq:gamma-theta}
\end{equation}
In what follows, we will not use the explicit form of
$F_{a'b'c'd'}$ given in eq.(\ref{eq:ads7-sol})
in order to trace the $F_{a'b'c'd'}$ dependence.
In Appendix \ref{sec:note}, we substitute the explicit
form of $F_{a'b'c'd'}$ into several quantities
which will appear in the following.
By using the relations
\begin{equation}
\mathcal{P}^{(\mathrm{LC})}_{\pm} \Gamma^{\bpu}
  = \Gamma^{\bpu} \mathcal{P}^{(\mathrm{LC})}_{\mp}~,
\quad
\mathcal{P}^{(\mathrm{LC})}_{\pm} \Gamma^{\bmi}
  = \Gamma^{\bmi} \mathcal{P}^{(\mathrm{LC})}_{\mp}~,
\quad
\mathcal{P}^{(\mathrm{LC})}_{\pm} \Gamma^{\hmp}
  = \Gamma^{\hmp} \mathcal{P}^{(\mathrm{LC})}_{\pm}~,
  \label{eq:P-gamma}
\end{equation}
we find that
\begin{equation}
\mathcal{M}^{2}[\Theta_{-}]
  =\mathcal{P}^{\mathrm{(LC)}}_{-}\mathcal{M}^{2}[\Theta_{-}]
   \mathcal{P}^{\mathrm{(LC)}}_{+}~,
 \label{eq:pm2p}
\end{equation}
and thus
\begin{equation}
\mathcal{M}^{4} [\Theta_{-}]=0~,
\quad
\bar{\Theta}_{-} \Gamma^{\hap} \mathcal{M}^{2}[\Theta_{-}]
=0~.
\label{eq:simplification1}
\end{equation}
It follows that in the gauge (\ref{eq:LCgauge1})
the supervielbein and the superconnection
become at most quartic order polynomials
of the fermionic coordinate $\Theta_{-}$. 
With the condition (\ref{eq:LCfermi}),
the pull-back of the covariant derivative onto the membrane
world-volume $\tilde{D}_{i} \Theta$ becomes
\begin{eqnarray}
\tilde{D}_{i} \Theta_{-} &\equiv&
  \partial_{i} \Theta_{-}
  + \partial_{i} X^{\hm}
    \left( \frac{1}{4}{\omega_{\hm}}^{\ha\hb} \Gamma_{\ha\hb}
            +e^{\ha}_{\hm} {T_{\ha}}^{\hb\hc\hd\he}
              F_{\hb\hc\hd\he} \right) \Theta_{-} \nonumber\\
&=&  \partial_{i} \Theta_{-}
     + \partial_{i} X^{+} e^{\bpu}_{+}  \Gamma^{\bmi}
       \left( \frac{1}{2L}\Gamma_{\br}
              +\frac{1}{288} 
                \Gamma^{b'c'd'e'} F_{b'c'd'e'} \right) \Theta_{-}
    \nonumber\\
&& \hspace{2.3em}
   {} + \partial_{i} X^{\hmp}\left(
      \frac{1}{4}{\omega_{\hmp}}^{\hap\hbp} \Gamma_{\hap\hbp}
       +e^{\hap}_{\hmp}
        {T_{\hap}}^{b'c'd'e'}F_{b'c'd'e'}
     \right) \Theta_{-}~.
   \label{eq:d-minus}
\end{eqnarray}
Here we have used the relation
${\omega_{+}}^{\bpu\br}=\frac{1}{L}e^{\bpu}_{+}$.

If we plugged the explicit form (\ref{eq:ads7-sol})
of $F_{a'b'c'd'}$ into eq.(\ref{eq:d-minus}), we would find
that $\tilde{D}_{i}$ contains
the projection operator\footnote{\ \ $\frac{1}{2L} \Gamma_{\br}
    +\frac{1}{288} \Gamma^{b'c'd'e'}F_{b'c'd'e'} 
    = \frac{1}{L} \Gamma_{\br}
       \mathcal{P}^{(\mathrm{M}5)}_{-}$~, $~~$
 $\frac{1}{4} {\omega_{\tm}}^{\ta \tb} \Gamma_{\ta \tb}
  +e^{\ta}_{\tm} {T_{\ta}}^{b'c'd'e'} F_{b'c'd'e'}
  = \frac{1}{L} e^{\ta}_{\tm} \Gamma_{\ta}
     \mathcal{P}^{(\mathrm{M}5)}_{-}$~.}
$\mathcal{P}^{(\mathrm{M}5)}_{-}$ \cite{LPR}\cite{Kallosh}\cite{PST}
defined as
\begin{equation}
\mathcal{P}^{(\mathrm{M}5)}_{\pm}
 =\frac{1 \pm \eta \Gamma_{\bar{r}}\hga}{2}~,
\quad
\hga = \frac{1}{4!} \epsilon^{a'b'c'd'}\Gamma_{a'b'c'd'}~.
\label{eq:M5-projector}
\end{equation}
We note that the projection operator
$\mathcal{P}^{(\mathrm{M}5)}_{\pm}$
commutes with $\mathcal{P}^{(\mathrm{LC})}_{+}$
and $\mathcal{P}^{(\mathrm{LC})}_{-}$.
We might further decompose the fermionic coordinate
$\Theta_{-}$ into two pieces as
\begin{equation}
\Theta_{-} = \lamm+ \psm~,
\quad
\Theta^{(\pm)}_{-}
 = \mathcal{P}^{(\mathrm{M}5)}_{\pm} \Theta_{-}~.
\label{eq:m5-projected}
\end{equation}
In Appendix \ref{sec:note},
we describe
several quantities in terms of
$\Theta^{(\pm)}_{-}$
after using the explicit form of $F_{a'b'c'd'}$.

Taking into account the bosonic condition in eq.(\ref{eq:LCgauge1})
as well as fermionic one, we obtain the following relations
 from eqs.(\ref{eq:pm2p}) and (\ref{eq:d-minus}),
\begin{eqnarray}
&&
\mathcal{M}^{2}[\Theta_{-}] \tilde{D}_{\acute{\imath}}
   \Theta_{-} =0~,
\quad \bar{\Theta}_{-} \Gamma^{\hap} \tilde{D}_{\acute{\imath}}=0~,
    \nonumber\\
&& \mathcal{M}^{2}[\Theta_{-}] \tilde{D}_{\tau}
\Theta_{-} = e_{+}^{\bpu}\mathcal{M}^{2}[\Theta_{-}]
\Gamma_{\bpu}
 \left(\frac{1}{2L} \Gamma_{\br}
       +\frac{1}{288} \Gamma^{a'b'c'd'}
        F_{a'b'c'd'}\right) \Theta_{-}~.
\end{eqnarray}

Gathering the relations obtained above,
we find that in the gauge (\ref{eq:LCgauge1})
the pull-back of the supervielbein onto the membrane world-volume
$\Pi^{A}_{i}$ becomes as follows:
the bosonic components $\Pi^{\ha}_{i}$ take the forms
\begin{eqnarray}
\Pi^{\bpu}_{\tau} &=& e^{\bpu}_{+}~,
\nonumber\\
\Pi^{\bmi}_{\tau} &=&
  \partial_{\tau} X^{-} e^{\bmi}_{-}
    -i\bar{\Theta}_{-} \Gamma^{\bmi} \partial_{\tau} \Theta_{-}
  \nonumber\\
  && {} -i\partial_{\tau} X^{\hmp} \bar{\Theta}_{-} \Gamma^{\bmi}
     \left( \frac{1}{4}{\omega_{\hmp}}^{\hap\hbp} \Gamma_{\hap\hbp}
        +e^{\hap}_{\hmp}{T_{\hap}}^{b'c'd'e'}
         F_{b'c'd'e'} \right) \Theta_{-}
  \nonumber\\
  &&{}-\frac{i}{12}  e^{\bpu}_{+}
       \bar{\Theta}_{-} \Gamma^{\bmi}
       \mathcal{M}^{2}[\Theta_{-}] \Gamma^{\bmi}
       \left(\frac{1}{2L} \Gamma_{\br}
             +\frac{1}{288}\Gamma^{b'c'd'e'}
              F_{b'c'd'e'} \right) \Theta_{-}~,
   \nonumber\\
\Pi^{\hap}_{\tau}
 &=& \partial_{\tau} X^{\hmp} e^{\hap}_{\hmp}
     -i e^{\bpu}_{+}
        \bar{\Theta}_{-} \Gamma^{\hap}\Gamma^{\bmi}
      \left(\frac{1}{2L} \Gamma_{\br}
             +\frac{1}{288} \Gamma^{b'c'd'e'}
              F_{b'c'd'e'} \right) \Theta_{-}~,
   \nonumber\\
\Pi^{\bpu}_{\acute{\imath}} &=& 0~, \nonumber\\
\Pi^{\bmi}_{\acute{\imath}} &=&
   \partial_{\acute{\imath}} X^{-} e_{-}^{\bmi}
    -i \bar{\Theta}_{-} \Gamma^{\bmi}
       \partial_{\acute{\imath}}\Theta_{-}
   \nonumber\\
   &&{}
 -i\partial_{\acute{\imath}}X^{\hmp}
     \bar{\Theta}_{-} \Gamma^{\bmi}
     \left(\frac{1}{4}{\omega_{\hmp}}^{\hap\hbp}\Gamma_{\hap\hbp}
          +e^{\hap}_{\hmp}
          {T_{\hap}}^{b'c'd'e'}F_{b'c'd'e'}
     \right)\Theta_{-}~, \nonumber\\
\Pi^{\hap}_{\acute{\imath}}
 &=& \partial_{\acute{\imath}}X^{\hmp} e_{\hmp}^{\hap}~,
 \label{eq:b-vielbein}
\end{eqnarray}
and the fermionic components
$\Pi_{i}^{\hat{\alpha}}$ take the forms
\begin{eqnarray}
\Pi_{\tau}&=& \partial_{\tau} \Theta_{-}
  + e^{\bpu}_{+}
   \left(1+\frac{1}{6}\mathcal{M}^{2}[\Theta_{-}]\right)
    \Gamma^{\bmi}
    \left(\frac{1}{2L}\Gamma_{\br}
         +\frac{1}{288} \Gamma^{a'b'c'd'}
         F_{a'b'c'd'} \right)\Theta_{-} \nonumber\\
 && {} + \partial_{\tau} X^{\hmp}
        \left(
        \frac{1}{4}{\omega_{\hmp}}^{\hap\hbp}\Gamma_{\hap\hbp}
          +e^{\hap}_{\hmp}
          {T_{\hap}}^{b'c'd'e'}F_{b'c'd'e'}
     \right)\Theta_{-}~,
 \nonumber\\
\Pi_{\acute{\imath}} &=& \partial_{\acute{\imath}}\Theta_{-}
 +\partial_{\acute{\imath}} X^{\hmp}
  \left(\frac{1}{4}{\omega_{\hmp}}^{\hap\hbp}\Gamma_{\hap\hbp}
          +e^{\hap}_{\hmp}
          {T_{\hap}}^{b'c'd'e'}F_{b'c'd'e'}
     \right)\Theta_{-}~.
  \label{eq:f-vielbein}
\end{eqnarray}
We thereby obtain the induced metric on the membrane world-volume,
\begin{eqnarray}
h_{\tau\tau}
&=& 2 \partial_{\tau} X^{-} G_{+-}
    + \partial_{\tau} X^{\hmp} \partial_{\tau} X^{\hnp} G_{\hmp\hnp}
    -i2 e^{\bpu}_{+} \bar{\Theta}_{-}\Gamma^{\bmi}
        \partial_{\tau} \Theta_{-}
    \nonumber\\
  &&{}-2i \partial_{\tau} X^{m'} e^{\bpu}_{+}
       \bar{\Theta}_{-}\Gamma^{\bmi}
        \left( \frac{1}{4} {\omega_{m'}}^{a'b'} \Gamma_{a'b'}
              -\frac{1}{2L}e^{a'}_{m'} \Gamma_{a'}\Gamma_{\br}
              -\frac{1}{24} e^{a'}_{m'} \Gamma^{b'c'd'}
                F_{a'b'c'd'} \right) \Theta_{-}
     \nonumber\\
   &&+\mathbb{F}~, \nonumber\\
u_{\acute{\imath}} &\equiv&
    h_{\tau\acute{\imath}} \nonumber\\
  &=& \partial_{\acute{\imath}} X^{-} G_{+-}
    +\partial_{\tau} X^{\hmp} \partial_{\acute{\imath}} X^{\hnp}
     G_{\hmp\hnp}
     -i e^{\bpu}_{+} \bar{\Theta}_{-} \Gamma^{\bmi}
        \partial_{\acute{\imath}} \Theta_{-} \nonumber\\
   && -i e^{\bpu}_{+} \partial_{\acute{\imath}} X^{m'}
       \bar{\Theta}_{-} \Gamma^{\bmi}
        \left(\frac{1}{4} {\omega_{m'}}^{a'b'}\Gamma_{a'b'}
               -\frac{1}{2L} e^{a'}_{m'} \Gamma_{a'}\Gamma_{\br}
               -\frac{1}{24} e^{a'}_{m'} \Gamma^{b'c'd'}
                 F_{a'b'c'd'} \right) \Theta_{-}~,
\nonumber\\
\bar{h}_{\acute{\imath}\acute{\jmath}}
&\equiv& h_{\acute{\imath}\acute{\jmath}}
= \partial_{\acute{\imath}}X^{\hmp} \partial_{\acute{\jmath}}X^{\hnp}
  G_{\hmp\hnp}~,
\end{eqnarray}
where $\mathbb{F}$ denotes four-Fermi terms
\begin{eqnarray}
\mathbb{F} &\equiv& -\frac{i}{6}  ( e^{\bpu}_{+} )^{2}
  \bar{\Theta}_{-} \Gamma^{\bmi} \mathcal{M}^{2} [\Theta_{-}]
  \Gamma^{\bmi}
  \left(\frac{1}{2L} \Gamma_{\br}
        +\frac{1}{288} \Gamma^{a'_{1}\cdots a'_{4}}
           F_{b'_{1}\cdots b'_{4}}
   \right) \Theta_{-}
   \nonumber\\
&& {}- ( e^{\bpu}_{+} )^{2}
   \left[ \bar{\Theta}_{-} \Gamma^{\bmi} \Gamma^{\hap}
        \left(\frac{1}{2L} \Gamma_{\br}
        +\frac{1}{288}  \Gamma^{b'_{1} \cdots b'_{4}}
           F_{c'_{1} \cdots c'_{4}}
   \right) \Theta_{-}\right] \times \nonumber\\
&& \hspace{5em} {}\times
    \left[\bar{\Theta}_{-} \Gamma^{\bmi} \Gamma_{\hap}
        \left(\frac{1}{2L} \Gamma_{\br}
        +\frac{1}{288} \Gamma^{c'_{1}\cdots c'_{4}}
           F_{c'_{1} \cdots c'_{4}}
   \right) \Theta_{-}\right]~.
   \label{eq:def-4-fermi}
\end{eqnarray}
{}Following the steps presented in
Appendix \ref{sec:derivation},
we may recast $\mathbb{F}$ into
\begin{eqnarray}
\mathbb{F}
&=& -(e^{\bpu}_{+})^{2} \left[
      \bar{\Theta}_{-} \Gamma^{\bmi} \Gamma^{\hap}
        \left(\frac{1}{2L} \Gamma_{\br}
         + \frac{1}{144} \Gamma^{b'_{1}\cdots b'_{4}}
            F_{b'_{1} \cdots b'_{4}} \right)
            \Theta_{-} \right]
       \times \nonumber\\
  && \hspace{5em} \times 
      \left[\bar{\Theta}_{-} \Gamma^{\bmi} \Gamma_{\hap}
        \left(\frac{1}{2L} \Gamma_{\br}
        +\frac{1}{288} \Gamma^{c'_{1}\cdots c'_{4}}
           F_{c'_{1} \cdots c'_{4}}
   \right) \Theta_{-}\right]
 \nonumber\\
&& {}+ \frac{1}{48} (e^{\bpu}_{+})^{2}
   \left( \bar{\Theta}_{-} \Gamma^{\bmi}\Gamma^{c'_{1}c'_{2}}
          \Theta_{-} \right) F_{c'_{1}\cdots c'_{4}}
          \times \nonumber\\
 && \hspace{6em} \times 
    \left[ \bar{\Theta}_{-} \Gamma^{\bmi} \Gamma^{c'_{3}c'_{4}}
      \left(\frac{1}{2L} \Gamma_{\br} + \frac{1}{288}
       \Gamma^{a'_{1}\cdots a'_{4}} F_{a'_{1} \cdots a'_{4}}
      \right) \Theta_{-} \right]~.
   \label{eq:4-fermi}
\end{eqnarray}

Substituting eqs.(\ref{eq:b-vielbein}) and (\ref{eq:f-vielbein})
into the lagrangian (\ref{eq:action1}),
we obtain
\begin{eqnarray}
\mathcal{L}&=& {} -T  \sqrt{\Delta \bar{h}}
   +T
  \frac{\epsilon^{\acute{\imath}\acute{\jmath}}}{2}
   \partial_{\tau}X^{m'_{1}}
   \partial_{\acute{\imath}}X^{m'_{2}}
   \partial_{\acute{\jmath}} X^{m'_{3}}
   C_{m'_{1}m'_{2}m'_{3}}
\nonumber\\
&& {}+ T\, \epsilon^{\acute{\imath}\acute{\jmath}}\,
       e^{\bpu}_{+}e^{\hap}_{\hmp}
       \partial_{\acute{\imath}} X^{\hmp}
       \;
   i \bar{\Theta}_{-} \Gamma^{\bmi} \Gamma_{\hap}
     \left[
      \partial_{\acute{\jmath}}\Theta_{-}
     + \partial_{\acute{\jmath}} X^{\hnp}
         \left( \frac{1}{4} {\omega_{\hnp}}^{\hbp\hcp}
                 \Gamma_{\hbp\hcp}+
     \right. \right.
     \nonumber\\
  && \hspace{9em} \left.\left.{}
                 +\frac{1}{4L} e^{\hbp}_{\hnp} 
                      \Gamma_{\hbp \br}
         +\frac{1}{192} e^{\hbp}_{\hnp}
             \Gamma^{c'_{1}\cdots c'_{4}}
             \Gamma_{\hbp}  F_{c'_{1}\cdots c'_{4}} \right)
         \Theta_{-}  \right]~,
    \label{eq:action2}
\end{eqnarray}
where
\begin{equation}
\Delta = -h_{\tau\tau} + u_{\acute{\imath}}
 \bar{h}^{\acute{\imath}\acute{\jmath}}u_{\acute{\jmath}}~,
 \quad \bar{h}^{\acute{\imath}\acute{\jmath}} h_{\acute{j}\acute{k}}
  =\delta^{\acute{\imath}}_{\acute{k}}~,
 \quad \bar{h} = \det \bar{h}_{\acute{\imath}\acute{\jmath}}~,
 \quad \epsilon^{\acute{\imath}\acute{\jmath}}
      = -\epsilon^{\tau \acute{\imath}\acute{\jmath}}~.
\end{equation}
Here we have used the relation (\ref{eq:gamma-symmetric}).
{}From the lagrangian (\ref{eq:action2}),
the canonical momenta are determined by
$$P_{-} = \frac{\partial \mathcal{L}}{\partial (\partial_{\tau}X^{-})}~,
\quad P_{\hmp}
  =\frac{\partial \mathcal{L}}{\partial (\partial_{\tau}X^{\hmp})}~,
\quad
\bar{P}_{\Theta_{-}}=
 \partial \mathcal{L} / \partial (\partial_{\tau} \Theta_{-})~.
$$
This yields
\begin{eqnarray}
&& P_{-}
= T \sqrt{\frac{\bar{h}}{\Delta}} G_{+-}~,
\qquad \quad
P_{\tm} 
 =T \sqrt{\frac{\bar{h}}{\Delta}} G_{\tm \tn}
  \left(\partial_{\tau} X^{\tn}
   -u_{\acute{\imath}} \bar{h}^{\acute{\imath}\acute{\jmath}}
    \partial_{\acute{\jmath}}X^{\tn} \right)~,
\nonumber\\
&& 
P_{r}=T \sqrt{\frac{\bar{h}}{\Delta}} G_{rr}
 \left(\partial_{\tau} r 
 - u_{\acute{\imath}}\bar{h}^{\acute{\imath}\acute{\jmath}}
   \partial_{\acute{\jmath}} r \right)~,
\nonumber\\
&& P_{m'} 
 =  T \sqrt{\frac{\bar{h}}{\Delta}}
      G_{m'n'} \left(\partial_{\tau} X^{n'}
     -u_{\acute{\imath}} \bar{h}^{\acute{\imath}\acute{\jmath}}
     \partial_{\acute{\jmath}} X^{n'}
     \right)
     +T \frac{\epsilon^{\acute{\imath}\acute{\jmath}}}{2}
         \partial_{\acute{\imath}} X^{n'_{1}}
         \partial_{\acute{\jmath}} X^{n'_{2}}
         C_{m'n'_{1}n'_{2}}
            \nonumber\\
&& \hspace{3em}
   {} -T \sqrt{\frac{\bar{h}}{\Delta}}
        i e^{\bpu}_{+} \bar{\Theta}_{-} \Gamma^{\bmi}
        \left( \frac{1}{4} {\omega_{m'}}^{a'b'} \Gamma_{a'b'}
              -\frac{1}{2L} e^{a'}_{m'} \Gamma_{a'} \Gamma_{\br}
              -\frac{1}{24} e^{a'}_{m'} \Gamma^{b'c'd'}
                  F_{a'b'c'd'} \right) \Theta_{-}~,
    \nonumber\\
&& \bar{P}_{\Theta_{-}}
    = -iT \sqrt{\frac{\bar{h}}{\Delta}}  e^{\bpu}_{+}
       \bar{\Theta_{-}} \Gamma^{\bmi}~.
   \label{eq:momenta}
\end{eqnarray}
The hamiltonian takes the form
\begin{eqnarray}
\mathcal{H}_{0}  &\equiv&  P_{-} \partial_{\tau} X^{-}
+ P_{\hmp} \partial_{\tau} X^{\hmp}
+\bar{P}_{\Theta_{-}} \partial_{\tau} \Theta_{-}
-\mathcal{L}
\label{eq:hamiltonian0} \\
&=& \frac{G_{+-}}{2P_{-}}
    \left[ T^{2} \bar{h}
        +P_{\tm}G^{\tm\tn}P_{\tn}
        +P_{r}G^{rr}P_{r} 
        +Q_{m'} G^{m'n'} Q_{n'} \right]
        \nonumber\\
 && {}-T\, \epsilon^{\acute{\imath}\acute{\jmath}}\,
       e^{\bpu}_{+}e^{\hap}_{\hmp}
       \partial_{\acute{\imath}} X^{\hmp}
    \; i\bar{\Theta}_{-} \Gamma^{\bmi} \Gamma_{\hap}
    \left[
      \partial_{\acute{\jmath}}\Theta_{-}
     + \partial_{\acute{\jmath}} X^{\hnp}
         \left( \frac{1}{4} {\omega_{\hnp}}^{\hbp\hcp}
                 \Gamma_{\hbp\hcp}
         \right. \right.
     \nonumber\\
 && \hspace{10em} \left.\left.
         {}+\frac{1}{4L} e^{\hbp}_{\hnp} 
                      \Gamma_{\hbp \br}
         +\frac{1}{192} e^{\hbp}_{\hnp}
             \Gamma^{c'_{1}\cdots c'_{4}}
             \Gamma_{\hbp}
                  F_{c'_{1}\cdots c'_{4}} \right)
         \Theta_{-}  \right]
      \nonumber\\
 && {}+\frac{1}{2} P_{-} \left[ \bar{\Theta}_{-} \Gamma^{\bmi}
       \Gamma^{\hap} \left( \frac{1}{2L} \Gamma_{\br}
        + \frac{1}{144} \Gamma^{b'_{1} \cdots b'_{4}}
           F_{b'_{1} \cdots b'_{4}} \right) \Theta_{-} \right] \times
      \nonumber\\
   && \hspace{5em} \times \left[
       \bar{\Theta}_{-} \Gamma^{\bmi} \Gamma_{\hap}
         \left(\frac{1}{2L} \Gamma_{\br} +
             \frac{1}{288} \Gamma^{c'_{1} \cdots c'_{4}}
             F_{c'_{1} \cdots c'_{4}} \right) \Theta_{-} \right]
      \nonumber\\
 && {} - \frac{1}{96} P_{-} \left( \bar{\Theta}_{-} \Gamma^{\bmi}
          \Gamma^{c'_{1}c'_{2}} \Theta_{-} \right)
          F_{c'_{1} \cdots c'_{4}}
          \left[ \bar{\Theta}_{-} \Gamma^{\bmi} \Gamma^{c'_{3}c'_{4}}
          \left( \frac{1}{2L} \Gamma_{\br}
                 + \frac{1}{288} \Gamma^{a'_{1} \cdots a'_{4}}
                    F_{a'_{1} \cdots a'_{4}} \right) \Theta_{-}
          \right],\nonumber
\end{eqnarray}
where
\begin{eqnarray}
Q_{m'} &\equiv&
     P_{m'}-\frac{T}{2} \epsilon^{\acute{\imath}\acute{\jmath}}
              \partial_{\acute{\imath}}X^{m'_{1}}
              \partial_{\acute{\jmath}}X^{m'_{2}}
              C_{m'm'_{1}m'_{2}} 
      \nonumber\\
  && {}+i \frac{P_{-}}{e^{\bmi}_{-}}
          \bar{\Theta}_{-} \Gamma^{\bmi}
           \left(\frac{1}{4} {\omega_{m'}}^{a'b'}\Gamma_{a'b'}
                 -\frac{1}{2L} e^{a'}_{m'} \Gamma_{a'}\Gamma_{\br}
                 -\frac{1}{24} e^{a'}_{m'} \Gamma^{b'c'd'}
                   F_{a'b'c'd'} \right) \Theta_{-}~.
  \label{eq:q}
\end{eqnarray}

Similarly to the flat case, the system has
primary constraints,
\begin{eqnarray}
\Phi_{\acute{\imath}} &\equiv&
  P_{-} \partial_{\acute{\imath}} X^{-}
  +P_{\hmp} \partial_{\acute{\imath}}X^{\hmp}
  +\bar{P}_{\Theta_{-}} \partial_{\acute{\imath}} \Theta_{-}
  \approx 0~,\nonumber\\
\bar{\chi} &\equiv&
  \bar{P}_{\Theta_{-}}
  +i\frac{P_{-}}{e^{\bmi}_{-}}  \bar{\Theta}_{-} \Gamma^{\bmi}
  \approx 0~.
  \label{eq:primary}
\end{eqnarray}
One can find that
the constraint $\Phi_{\acute{\imath}}$
is first class
and the constraint $\bar{\chi}$ is  second class.
{}Following the prescription for the constrained hamiltonian
systems \cite{Dirac}\cite{HRT},
we introduce the total hamiltonian
\begin{equation}
\mathcal{H}_{\mathrm{T}} =  \mathcal{H}_{0}
  +c^{\acute{\imath}} \Phi_{\acute{\imath}}~,
  \label{eq:totalhamiltonian}
\end{equation}
where $c^{\acute{\imath}}$ is a Lagrange multiplier.
Since $\bar{\chi}$ is a second class constraint,
we solve this constraint
and hence have not added this constraint
in the total hamiltonian (\ref{eq:totalhamiltonian}).
We find that
no further constraints emerge from the consistency conditions
of the constraints (\ref{eq:primary}),
i.e., there are no secondary constraints
in this system.

In order to fix the remaining gauge symmetry generated by the
first class constraint $\Phi_{\acute{\imath}}$,
we further impose conditions.
Now we look for the gauge in which the dynamics of $P_{-}(\tau,\sigma)$
becomes trivial, $\partial_{\tau} P_{-}=0$.
Since the hamiltonian $\mathcal{H}_{\mathrm{T}}$ is independent
of the coordinate $X^{-}$ conjugate to $P_{-}$
except for the constraint term\footnote{Because
of this fact, the field $X^{-}$ can be eliminated
{}from the resulting hamiltonian \cite{dWPP-curve}.
This originates from the fact
that all the components of the metric $G_{\hm\hn}$
and the 3-form potential $C_{\hm\hn\hp}$ are independent
of the coordinate $X^{-}$ and
 the components $C_{+-\hm}$ and $C_{-\hm\hn}$ of the
3-form potential are vanishing in this background.}
$\Phi_{\acute{\imath}}c^{\acute{\imath}}$,
we can achieve such a gauge, in the same way as
the flat case \cite{Hoppe}\cite{BS-Tanii}\cite{dWHN},
by imposing 
\begin{equation}
u_{\acute{\imath}}\approx 0~.
\label{eq:LCgauge2}
\end{equation}
We will henceforth refer to the gauge defined by eqs.(\ref{eq:LCgauge1})
and (\ref{eq:LCgauge2}) as the light-cone gauge.
The condition (\ref{eq:LCgauge2}) leads to
$c^{\acute{\imath}}\approx 0$
and thus the resulting hamiltonian becomes independent of $X^{-}$.
This leads to $\partial_{\tau} P_{-}=0$ and
we may set
\begin{equation}
P_{-}(\sigma) = P_{-}^{(0)}\sqrt{w(\sigma)}~,
\label{eq:scalardensity}
\end{equation}
where $\sqrt{w(\sigma)}$ is a scalar density in the spatial
directions of the membrane world-volume, $\Sigma_{(2)}$.
Now that we are considering a closed supermembrane,
$\Sigma_{(2)}$ is a compact space.
We normalize the area of $\Sigma_{(2)}$
as $\int d^{2}  \sigma \sqrt{w(\sigma)}=1$.
$P_{-}^{(0)}$ is therefore the zero mode of $P_{-}(\sigma)$:
$P_{-}^{(0)} = \int d^{2}\sigma P_{-}(\sigma)$.
Since $c^{\acute{\imath}}\approx 0$,
the hamiltonian $\mathcal{H}_{\mathrm{T}}$ introduced
in eq.(\ref{eq:totalhamiltonian}) turns out to take the same
form as $\mathcal{H}_{0}$ in eq.(\ref{eq:hamiltonian0})
with the identification (\ref{eq:scalardensity}).

It follows from the condition (\ref{eq:LCgauge2})
that $X^{-}$ is expressed in terms of other fields as
\begin{eqnarray}
\partial_{\acute{\imath}} X^{-}
 &=& -\frac{1}{G_{+-}}
       \partial_{\tau} X^{\hmp}
       \partial_{\acute{\imath}}X^{\hnp}
        G_{\hmp\hnp}
     +i \frac{1}{e^{\bmi}_{-}} \bar{\Theta}_{-} \Gamma^{\bmi}
        \partial_{\acute{\imath}} \Theta_{-} 
           \nonumber\\
  && {} +i \frac{1}{e^{\bmi}_{-}} \bar{\Theta}_{-}
        \Gamma^{\bmi} \left(\frac{1}{4} {\omega_{m'}}^{a'b'}\Gamma_{a'b'}
           -\frac{1}{2L} e_{m'}^{a'} \Gamma_{a'}\Gamma_{\br}
           -\frac{1}{24} e^{a'}_{m'} \Gamma^{b'c'd'} F_{a'b'c'd'}\right)
          \Theta_{-} \partial_{\acute{\imath}} X^{m'}
         \nonumber\\
 &=&   -\frac{1}{ P^{(0)}_{-} \sqrt{w} }
  P_{\hmp} \partial_{\acute{\imath}} X^{\hmp}
  +i\frac{1}{e^{\bmi}_{-}}
  \bar{\Theta}_{-} \Gamma^{\bmi} \partial_{\acute{\imath}}
  \Theta_{-}~.
   \label{eq:xminus}
\end{eqnarray}
The fields $X^{-}(\tau,\sigma)$ and $P_{-}(\tau,\sigma)$
are thus no longer independent
physical degrees of freedom except for their zero modes
$q^{-}(\tau)$ and $P^{(0)}_{-}$, where $q^{-}(\tau)$ is defined
as
\begin{equation}
q^{-}(\tau) = \int d^{2}\sigma
   \sqrt{w(\sigma)} X^{-}(\tau,\sigma)~.
\end{equation}
{}From the original canonical commutation relation
$(X^{-}(\tau,\sigma),P_{-}(\tau,\sigma'))_{\mathrm{P}}
  = \delta^{(2)} (\sigma,\sigma')$,
where $\delta^{(2)}(\sigma,\sigma')$ denotes the delta function
on $\Sigma_{(2)}$,
one can find that $q^{-}(\tau)$
and $P_{-}^{(0)}$ obey the commutation relation
\begin{equation}
\left(q^{-}(\tau), P_{-}^{(0)}\right)_{\mathrm{P}}=1~.
\end{equation}

Now we make a brief comment on the relation between
$h_{\tau\tau}$ and $\bar{h}$ in this gauge.
Combined with the definition of $P_{-}$ in eq.(\ref{eq:momenta}),
the relation (\ref{eq:scalardensity}) yields
\begin{equation}
h_{\tau\tau} = - \left(\frac{T}{\sqrt{w(\sigma)} P^{(0)}_{-}}
\right)^{2} G_{+-} \bar{h}~.
\end{equation}
We note that
the field dependent factor $G_{+-}=2r/L$  appears
in the proportional coefficient between $h_{\tau\tau}$
and $\bar{h}$.
This implies that,
unlike the flat case, we cannot accomplish
the `conformally-flat-like' world-volume metric
 in the $AdS_{7} \times S^{4}$ case.
The situation is quite similar to the light-cone superstring
on $AdS_{5} \times S^{5}$ \cite{MT-LC},
in which conformally flat world-sheet metric is not
allowed because it is incompatible with the
equations of motion for the string coordinate $X^{+}$ in the
light-cone gauge \cite{MT-LC}.

Using eq.(\ref{eq:xminus}), one can find that
\begin{equation}
\Phi_{\acute{\imath}} = \bar{\chi} \partial_{\acute{\imath}}
   \Theta_{-}~.
\end{equation}
This implies that the constraint $\Phi_{\acute{\imath}}$
is reduced to the second class constraint
$\bar{\chi}$ and need not be considered
in the light-cone gauge.
By using the second class constraint $\bar{\chi}$,
we evaluate the Dirac brackets among the canonical variables,
\begin{eqnarray}
 \left(X^{\hmp}(\tau,\sigma), P_{\hnp}(\tau,\sigma')
   \right)_{\mathrm{DB}}
&=& \delta^{\hmp}_{\hnp} \delta^{(2)}(\sigma,\sigma')~,
\nonumber\\
 \left( q^{-}(\tau), P_{-}^{(0)} \right)_{\mathrm{DB}}
  &=& 1~, \nonumber\\
\left(\Theta_{-}^{\hat{\alpha}}(\tau,\sigma),
    \bar{P}_{\Theta_{-}\hat{\beta}}(\tau,\sigma')
    \right)_{\mathrm{DB}}
   &=& \frac{1}{2}
   {( \mathcal{P}^{(\mathrm{LC})}_{-})^{\hat{\alpha}}}_{\hat{\beta}}
   \delta^{(2)}(\sigma,\sigma')~,
  \nonumber\\
\left( q^{-}(\tau),\Theta_{-}^{\hat{\alpha}}(\tau,\sigma)
\right)_{\mathrm{DB}}
&=& -\frac{1}{2P^{(0)}_{-}} \Theta^{\hat{\alpha}}_{-}(\tau,\sigma)~,
\nonumber\\
\left( P_{r}(\tau,\sigma), \Theta^{\hat{\alpha}}_{-}(\tau,\sigma')
\right)_{\mathrm{DB}}
&=& -\frac{1}{4r(\tau,\sigma)} \Theta^{\hat{\alpha}}_{-}(\tau,\sigma)
  \delta^{(2)}(\sigma,\sigma')~.
  \label{eq:DB1}
\end{eqnarray}
The third relation in the above
is also expressed in the following way,
\begin{equation}
\left( \Theta^{\hat{\alpha}}_{-}(\tau,\sigma),
       \Theta^{\hat{\beta}}_{-}(\tau,\sigma')
\right)_{\mathrm{DB}}
=- \frac{i}{4P^{(0)}_{-}} \frac{1}{\sqrt{w(\sigma)}}
   e^{\bmi}_{-}\left( \mathcal{P}^{\mathrm{(LC)}}_{-} \Gamma^{\bpu}
                       \mathcal{C} \right)^{\hat{\alpha}\hat{\beta}}
   \delta^{(2)}(\sigma,\sigma')~.
   \label{eq:DB2}
\end{equation}
We note that the fermionic coordinate $\Theta_{-}(\tau,\sigma)$
does not commute with the bosonic canonical variables
$q^{-}(\tau)$ and $P_{r}(\tau,\sigma)$.

In a similar way
to the flat case \cite{Hoppe}\cite{BS-Tanii}\cite{dWHN},
there still exists a constraint, i.e.\ the integrability
condition for $\partial_{\acute{\imath}}X^{-}$
given in eq.(\ref{eq:xminus}).
The integrability condition is locally described by
$P^{(0)}_{-}\epsilon^{\acute{\imath}\acute{\jmath}}
\partial_{\acute{\imath}} \partial_{\acute{\jmath}} X^{-}
/\sqrt{w}=0$, which reads
\begin{eqnarray}
\varphi &\equiv&
{} - P_{-}^{(0)}
          \left\{
          \frac{1}{G_{+-}} \partial_{\tau}X^{\hmp}
                 G_{\hmp\hnp}\, , \,
           X^{\hnp} \right\}
         +i P_{-}^{(0)}\left\{
             \frac{1}{e^{\bmi}_{-}} \bar{\Theta}_{-}\, , \,
              \Gamma^{\bmi} \Theta_{-}
    \right\} \nonumber\\
&&  {}+i P_{-}^{(0)}
     \left\{
       \frac{1}{e^{\bmi}_{-}} \bar{\Theta}_{-} \Gamma^{\bmi}
       \left( \frac{1}{4}{\omega_{m'}}^{a'b'} \Gamma_{a'b'}
              -\frac{1}{2L} e^{a'}_{m'}\Gamma_{a'}\Gamma_{\br}
              -\frac{1}{24} e^{a'}_{m'}\Gamma^{b'c'd'}
                 F_{a'b'c'd'}
        \right) \Theta_{-} \, , \,
      X^{m'} \right\}
   \nonumber\\
&=& {} - \left\{
  \frac{P_{\hmp}}{\sqrt{w}} \, , \,
   X^{\hmp}  \right\}
  +i P^{(0)}_{-}\left\{
    \frac{1}{e^{\bmi}_{-}} \bar{\Theta}_{-} \, , \,
    \Gamma^{\bmi} \Theta_{-} \right\}
    \approx 0~,
    \label{eq:apdgenerator}
\end{eqnarray}
where the bracket $\{ \ast\, , \, \ast \}$ is defined
as
\begin{equation}
\{ A\, , \, B \} =
\frac{\epsilon^{\acute{\imath}\acute{\jmath}}}{\sqrt{w}}
\partial_{\acute{\imath}} A \partial_{\acute{\jmath}} B
\end{equation}
for arbitrary functions $A$ and $B$ on $\Sigma_{(2)}$.
By using the Dirac brackets (\ref{eq:DB1}) and (\ref{eq:DB2}),
we can show that the constraint $\varphi$ generates
area-preserving diffeomorphisms (APD),
\begin{eqnarray}
&& \int d^{2} \sigma' \sqrt{w(\sigma')} \,f(\sigma')
    \left( X^{\hmp}(\sigma), \varphi(\sigma') \right)_{\mathrm{DB}}
    = \left\{ f(\sigma)\, , \, X^{\hmp}(\sigma) \right\}~,
    \nonumber\\
&& \int d^{2} \sigma' \sqrt{w(\sigma')} \, f(\sigma')
 \left( \Theta^{\hat{\alpha}}_{-}(\sigma), \varphi(\sigma')
 \right)_{\mathrm{DB}}
 =\left\{ f(\sigma)\, , \, \Theta^{\hat{\alpha}}_{-}(\sigma)\right\}~,
\end{eqnarray}
where $f(\sigma)$ is an arbitrary function on $\Sigma_{(2)}$.
While we might impose a further condition to fix this residual
APD gauge symmetry, we leave it unfixed,
following the treatment of \cite{dWHN}
to obtain quantum mechanics.
When $\Sigma_{(2)}$ is a Riemann surface of genus $g$,
there exist $2g$ global integrability conditions
(see e.g.\ \cite{dWMN})
in addition to the local one (\ref{eq:apdgenerator}).
These constraints generate APD transformations as well.

We can construct the lagrangian $\mathcal{L}_{\mathrm{APD}}$
which reproduces the light-cone gauge hamiltonian
through the Legendre transformation,
$\mathcal{L}_{\mathrm{APD}}
 = P_{\hmp} \partial_{\tau} X^{\hmp}
   +\bar{P}_{\Theta_{-}} \partial_{\tau} \Theta_{-}
   -\mathcal{H}_{\mathrm{T}}$.
We find that $\mathcal{L}_{\mathrm{APD}}$ takes
the form
\begin{eqnarray}
\lefteqn{
\frac{1}{\sqrt{w}} \mathcal{L}_{\mathrm{APD}}}
\label{eq:apdlagrangian1} \\
 &=&  \frac{P^{(0)}_{-} G_{\hmp \hnp} }{2 G_{+-}}
      \mathcal{D}_{\tau} X^{\hmp}
                   \mathcal{D}_{\tau} X^{\hnp}
  -\frac{G_{+-}}{4P^{(0)}_{-}} T^{2}
       \{ X^{\hmp},X^{\hnp}\}
       \{ X^{\hpp},X^{\hqp}\} G_{\hmp \hpp} G_{\hnp \hqp}
         \nonumber\\
  &&  {}+ \frac{T}{2} C_{m'_{1}m'_{2}m'_{3}} 
         \mathcal{D}_{\tau} X^{m'_{1}} \{ X^{m'_{2}},X^{m'_{3}}\}
     \nonumber\\
  && {}-\frac{P^{(0)}_{-}}{e^{\bmi}_{-}}
      i \bar{\Theta}_{-} \Gamma^{\bmi} \mathcal{D}_{\tau}
         \Theta_{-}
     +T e^{\bpu}_{+} i\bar{\Theta}_{-} \Gamma^{\bmi}
      \Gamma_{\hap} e^{\hap}_{\hmp}
       \{ X^{\hmp},\Theta_{-} \}
     \nonumber\\
   && {}-\frac{P^{(0)}_{-}}{e^{\bmi}_{-}}
        \mathcal{D}_{\tau} X^{m'}
        i \bar{\Theta}_{-} \Gamma^{\bmi}
        \left(\frac{1}{4}{\omega_{m'}}^{a'b'} \Gamma_{a'b'}
            -\frac{1}{2L} e^{a'}_{m'} \Gamma_{a'}\Gamma_{\br}
            -\frac{1}{24} e^{a'}_{m'} \Gamma^{b'c'd'}
              F_{a'b'c'd'} \right) \Theta_{-}
       \nonumber\\
   && {}+T  e^{\bpu}_{+} e^{\hap}_{\hmp}
         \{ X^{\hmp},X^{\hnp} \} \times
        \nonumber\\
    && \quad {}\times i
       \bar{\Theta}_{-} \Gamma^{\bmi} \Gamma_{\hap}
       \left( \frac{1}{4} {\omega_{\hnp}}^{\hbp \hcp} \Gamma_{\hbp\hcp}
             +\frac{1}{4L} e^{\hbp}_{\hnp} \Gamma_{\hbp\br}
             +\frac{1}{192} e^{\hbp}_{\hnp}
               \Gamma^{c'_{1} \cdots c'_{4}} \Gamma_{\hbp}
                F_{c'_{1}\cdots c'_{4}} 
       \right)\Theta_{-}
        \nonumber\\
   && {}-\frac{1}{2} P^{(0)}_{-} \left[ \bar{\Theta}_{-} \Gamma^{\bmi}
       \Gamma^{\hap} \left( \frac{1}{2L} \Gamma_{\br}
        + \frac{1}{144} \Gamma^{b'_{1} \cdots b'_{4}}
           F_{b'_{1} \cdots b'_{4}} \right) \Theta_{-} \right] \times
      \nonumber\\
   && \hspace{5em} \times \left[
       \bar{\Theta}_{-} \Gamma^{\bmi} \Gamma_{\hap}
         \left(\frac{1}{2L} \Gamma_{\br} +
             \frac{1}{288} \Gamma^{c'_{1} \cdots c'_{4}}
             F_{c'_{1} \cdots c'_{4}} \right) \Theta_{-} \right]
      \nonumber\\
 && {} + \frac{1}{96} P^{(0)}_{-} \left( \bar{\Theta}_{-} \Gamma^{\bmi}
          \Gamma^{c'_{1}c'_{2}} \Theta_{-} \right)
          F_{c'_{1} \cdots c'_{4}}
          \left[ \bar{\Theta}_{-} \Gamma^{\bmi} \Gamma^{c'_{3}c'_{4}}
          \left( \frac{1}{2L} \Gamma_{\br}
                 + \frac{1}{288} \Gamma^{a'_{1} \cdots a'_{4}}
                    F_{a'_{1} \cdots a'_{4}} \right) \Theta_{-}
          \right]. \nonumber
\end{eqnarray}
Here we have introduced the APD gauge field $v(\tau, \sigma)$
and replaced the $\tau$-derivative $\partial_{\tau}$
with the covariant derivative $\mathcal{D}_{\tau}$
defined  \cite{dWHN} as
\begin{equation}
\mathcal{D}_{\tau} X^{\hmp}
 = \partial_{\tau} X^{\hmp} -\{ v, X^{\hmp} \}~,
\quad
\mathcal{D}_{\tau} \Theta_{-}
 = \partial_{\tau} \Theta_{-} - \{ v, \Theta_{-} \}~.
\end{equation}
The hamiltonian $\mathcal{H}_{\mathrm{T}}$
can be interpreted as that
in the `temporal gauge', $v=0$.
The constraint $\varphi=0$ may be obtained as the Gauss-law
constraint
in the temporal gauge:
\begin{equation}
0=
\left. \frac{\delta S_{\mathrm{APD}}}{\delta v} \right|_{v=0}
= - \sqrt{w} \varphi~,
\end{equation}
where $S_{\mathrm{APD}} = \int d^{3}\xi \mathcal{L}_{\mathrm{APD}}$.

The lagrangian (\ref{eq:apdlagrangian1})
does not explicitly depend on the geometry of $\Sigma_{(2)}$.
This allows us to
reinterpret the spatial directions of the membrane world-volume
$\Sigma_{(2)}$ as an internal space on which the APD gauge
transformations act
and regard the lagrangian
\begin{equation}
L_{\mathrm{APD}}=\int d^{2}\sigma \mathcal{L}_{\mathrm{APD}}
\label{eq:apdlagrangian2}
\end{equation}
as that of quantum mechanics.
We thus obtain the quantum mechanical system
with APD gauge symmetry which describes the supermembrane
on $AdS_{7} \times S^{4}$.

\subsection{Field redefinition and 
$SO(1,1)\times SO(9)$ decomposition in the fermionic sector}
\label{sec:redefinition}
We remark on the field-dependent
rescaling of the fermionic coordinate
$\Theta_{-}$,
\begin{equation}
\Theta_{-} \longmapsto \tilde{\Theta}_{-}
  = \mathcal{F} \left[X^{\hmp};P_{-}^{(0)} \right] \Theta_{-}~,
 \label{eq:rescaling}
\end{equation}
where $\mathcal{F}\left[ X^{\hmp};P_{-}^{(0)} \right]$ is
an arbitrary scalar-valued functional
of $X^{\hmp}$ and $P^{(0)}_{-}$.
By using eq.(\ref{eq:gamma-symmetric}),
one can readily show that this transformation does not generate
in the lagrangian (\ref{eq:apdlagrangian2}) a new term
which depends on the derivatives $\partial_{i} X^{\hmp}$.
It follows that, through the field redefinition
(\ref{eq:rescaling}), we may change the 
normalization
of the fermionic coordinate $\Theta_{-}$ in the lagrangian
(\ref{eq:apdlagrangian2}) even in a field-dependent way
 without altering the momenta $P_{\hmp}$.

For later convenience, we decompose the $SO(10,1)$ gamma matrices
$\Gamma^{\ha}$ into $SO(1,1)\times SO(9)$ gamma matrices
following the manipulation
given in Appendix \ref{sec:gamma} and describe
several formulae obtained in the last subsection in terms of
the $SO(9)$ spinors and the gamma matrices $\gamma_{\hap}$.
Eq.(\ref{eq:halfspinor}) enables us to express
the $SO(10,1)$ Majorana spinor
$\Theta_{-}^{\hat{\alpha}}$
in terms of the $SO(9)$ Majorana spinor $\theta^{\alpha}$
as
\begin{equation}
\Theta_{-} = \frac{1}{2^{\frac{3}{4}} \sqrt{P_{-}^{(0)}} }
     \left( \frac{2r}{L} \right)^{\frac{1}{4}}\,
   \left( \begin{array}{c}
            0 \\ \theta
           \end{array} \right)~.
   \label{eq:redefinition}
\end{equation}
Here we have made  the field redefinition
(\ref{eq:rescaling}) with the rescaling factor
$\mathcal{F} =2^{\frac{3}{4}} \sqrt{P_{-}^{(0)}} 
     \left( \frac{L}{2r} \right)^{\frac{1}{4}}$ as well,
in order that
the new fermionic coordinate $\theta$ should commute with
the bosonic canonical variables.
In fact, the commutation relations (\ref{eq:DB1}) and (\ref{eq:DB2})
involving the fermionic coordinate are modified into
\begin{eqnarray}
\left( \theta^{\alpha}(\tau,\sigma), \theta^{\beta}(\tau,\sigma')
\right)_{\mathrm{DB}}
&=& -i \delta^{\alpha\beta} \frac{1}{\sqrt{w}}
    \delta^{(2)}(\sigma,\sigma')~,
  \nonumber\\
\left( q^{-}(\tau), \theta^{\alpha}(\tau,\sigma)\right)_{\mathrm{DB}}
&=& \left(P_{r}(\tau,\sigma), \theta^{\alpha}(\tau,\sigma')
    \right)_{\mathrm{DB}}=0~.  
\end{eqnarray}

In terms of $\theta$, eq.(\ref{eq:xminus})
is  described by
\begin{eqnarray}
\partial_{\acute{\imath}} X^{-}
  &=& {}- \frac{L}{2r} \partial_{\tau} X^{\hmp}
       \partial_{\acute{\imath}} X^{hnp} G_{\hmp \hnp}
      -\frac{i}{2P_{-}^{(0)}} \theta \partial_{\acute{\imath}} \theta
   \nonumber\\
  && {}-\frac{i}{2P^{(0)}_{-}} \theta \left(
     \frac{1}{4} {\omega_{m'}}^{a'b'} \gamma_{a'b'}
     -\frac{1}{2L} e^{a'}_{m'} \gamma_{a'} \gamma_{\br}
     +\frac{1}{24} e^{a'}_{m'}\gamma^{b'c'd'} F_{a'b'c'd'}
     \right) \theta \partial_{\acute{\imath}} X^{m'}
   \nonumber\\
   &=& {}-\frac{1}{P^{(0)}_{-}} \left(
        \frac{P_{\hmp}}{\sqrt{w}} \partial_{\acute{\imath}} X^{\hmp}
            +\frac{i}{2} \theta \partial_{\acute{\imath}} \theta
         \right)~.
\end{eqnarray}
The APD constraint $\varphi$ takes the form
\begin{eqnarray}
\varphi &=& {}-\frac{P^{(0)}_{-}}{2}
  \left\{ \frac{L}{2r} \partial_{\tau} X^{\hmp} G_{\hmp\hnp}
          \, , \,    X^{\hnp} \right\}
   -\frac{i}{2} \left\{ \theta, \theta \right\} \nonumber\\
 && {}-\frac{i}{2} \left\{
      \theta \left( \frac{1}{4} {\omega_{m'}}^{a'b'} \gamma_{a'b'}
      -\frac{1}{2L} e^{a'}_{m'} \gamma_{a'} \gamma_{\br}
      +\frac{1}{24} e^{a'}_{m'} \gamma^{b'c'd'} F_{a'b'c'd'}
      \right) \theta\, , \,
      X^{m'} \right\} \nonumber\\
  &=& - \left\{ \frac{P_{\hmp}}{\sqrt{w}}\, ,\, X^{\hmp} \right\}
      -\frac{i}{2} \left\{ \theta\, , \, \theta \right\}~.
\end{eqnarray}
The lagrangian $L_{\mathrm{APD}}$
is expressed as
\begin{eqnarray}
&& \! \! 
L_{\mathrm{APD}}=\int d^{2}\sigma \sqrt{w(\sigma)}\times
\nonumber\\
 && {}\times \left[
      {} \frac{P^{(0)}_{-} }{2}
       \frac{L}{2r}  G_{\hmp \hnp} 
      \mathcal{D}_{\tau} X^{\hmp}
                   \mathcal{D}_{\tau} X^{\hnp}
  -\frac{T^{2}}{4P^{(0)}_{-}} \frac{2r}{L}
       \{ X^{\hmp},X^{\hnp}\}
       \{ X^{\hpp},X^{\hqp}\} G_{\hmp \hpp} G_{\hnp \hqp}
   \right.   \nonumber\\
  && \quad {}+ \frac{T}{2} C_{m'_{1}m'_{2}m'_{3}} 
         \mathcal{D}_{\tau} X^{m'_{1}} \{ X^{m'_{2}},X^{m'_{3}}\}
 \nonumber\\
 &&\quad {}  +\frac{i}{2} \theta \mathcal{D}_{\tau} \theta
      +\frac{T}{2P^{(0)}_{-}}
         \frac{2r}{L} i \theta
       \gamma_{\hap} e^{\hap}_{\hmp} \{ X^{\hmp}, \theta\}
     \nonumber\\
&&\quad {}+\frac{i}{2} \mathcal{D}_{\tau} X^{m'}
         \theta \left(\frac{1}{4} {\omega_{m'}}^{a'b'} \gamma_{a'b'}
             -\frac{1}{2L} e^{a'}_{m'} \gamma_{a'} \gamma_{\br}
             +\frac{1}{24} e^{a'}_{m'} \gamma^{b'c'd'} F_{a'b'c'd}
          \right)\theta \nonumber\\
&& \quad {}+  \frac{T}{2P^{(0)}_{-}} \frac{2r}{L} e^{\hap}_{\hmp}
        \{ X^{\hmp}, X^{\hnp} \}
        \times \nonumber\\
&& \quad \qquad  \times {} i
    \theta \gamma_{\hap}
      \left( \frac{1}{4} {\omega_{\hnp}}^{\hbp\hcp} \gamma_{\hbp\hcp}
             +\frac{1}{4L} e^{\hbp}_{\hnp} \gamma_{\hbp\br}
             -\frac{1}{192} e^{\hbp}_{\hnp}
              \gamma^{c'_{1} \cdots c'_{4}}
              \gamma_{\hbp}
               F_{c'_{1}\cdots c'_{4}} \right) \theta
      \nonumber\\
 && \quad {}- \frac{1}{8P^{(0)}_{-}} \frac{2r}{L}
      \left[ \theta \gamma^{\hap}
             \left(\frac{1}{2L} \gamma_{\br} 
             - \frac{1}{144} \gamma^{b'_{1} \cdots b'_{4}}
               F_{b'_{1} \cdots b'_{4}} \right) \theta \right]
       \! \left[ \theta \gamma_{\hap} \left(
             \frac{1}{2L} \gamma_{\br}
             - \frac{1}{288} \gamma^{c'_{1} \cdots c'_{4}}
               F_{c'_{1} \cdots c'_{4}} \right) \theta \right]
     \nonumber\\
 &&  \quad  \left. {} - \frac{1}{384 P^{(0)}_{-}}
      \frac{2r}{L} \left( \theta \gamma^{c'_{1}c'_{2}} \theta \right)
          F_{c'_{1} \cdots c'_{4}}
          \left[ \theta \gamma^{c'_{3}c'_{4}}
                \left( \frac{1}{2L} \gamma_{\br}
                       -\frac{1}{288} \gamma^{a'_{1} \cdots a'_{4}}
                       F_{a'_{1} \cdots a'_{4}} \right) \theta
           \right]  \right]~.
    \label{eq:apdlagrangian5}
\end{eqnarray}

\subsection{Matrix regularization}
The algebra generated by APD is approximated by the large $N$ $u(N)$
Lie algebra.
This provides a prescription to regularize the two-dimensional
continuum `internal space' $\Sigma_{(2)}$
by the $N^{2}$-dimensional vector space of the adjoint
representation of the $U(N)$ group.
Following the standard prescription
\cite{Hoppe}\cite{dWHN}\cite{FFZ},
we replace the fields
on the membrane-world volume with $\tau$-dependent
$N\times N$ hermitian matrices,
the APD gauge field $v$ with the $U(N)$ gauge field $A_{\tau}$,
the integral over $\Sigma_{(2)}$ with the matrix trace
and the bracket $\{\ast,\ast\}$ with the matrix
commutator $[\ast,\ast]$:
\begin{eqnarray}
X^{\hmp}(\tau,\sigma)
   \; \stackrel{N\rightarrow \infty}{\longleftarrow}\;
   {X^{\hmp}(\tau)^{I}}_{J}~, &&
\theta^{\alpha}(\tau,\sigma)
   \; \stackrel{N\rightarrow \infty}{\longleftarrow}
   \;   {\theta^{\alpha}(\tau)^{I}}_{J}~, \nonumber\\
v(\tau,\sigma)
   \;\stackrel{N\rightarrow \infty}{\longleftarrow}
   \;   \frac{1}{2\pi N}{A_{\tau}(\tau)^{I}}_{J}~, &&
\frac{P_{\hmp}(\tau,\sigma)}{\sqrt{w(\sigma)}}
   \;\stackrel{N\rightarrow \infty}{\longleftarrow}
   \; N {P_{\hmp}(\tau)^{I}}_{J}~,
   \nonumber\\
{}\{\ast, \ast\}
   \;\stackrel{N\rightarrow \infty}{\longleftarrow}
   \; -i2\pi N [ \ast, \ast]~,&&
\int d^{2} \sigma \,\sqrt{w(\sigma)}
   \;\stackrel{N\rightarrow \infty}{\longleftarrow}
   \; \frac{1}{N} \mathrm{Tr}~,
 \label{eq:matrix-reg}
\end{eqnarray}
where $I,J=1,\ldots,N$ denote the matrix indices.
Applying this prescription to the lagrangian (\ref{eq:apdlagrangian5}),
we obtain the action of the matrix model for the supermembrane
on $AdS_{7} \times S^{4}$,
\begin{eqnarray}
&& S_{\mathrm{matrix}}
  = S_{\mathrm{matrix}}^{(B)}
   +S_{\mathrm{matrix}}^{(F2)}
   +S_{\mathrm{matrix}}^{(F4)}~, \nonumber\\
&& \quad {} S^{(B)}_{\mathrm{matrix}}
= \int d\tau \,\frac{P_{-}^{(0)}}{N} \mathrm{Tr} \left[
   \frac{1}{2} \frac{L}{2r}
     \mathcal{D}_{\tau} X^{\hmp}
     \mathcal{D}_{\tau} X^{\hnp} G_{\hmp\hnp}
   \right. \nonumber\\
&& \hspace{11em} {}
   +\frac{1}{4} \left(
      \frac{N}{P_{-}^{(0)}} \frac{1}{2\pi l_{p}^{3}} \right)^{2}
      \frac{2r}{L}  [X^{\hmp},X^{\hnp}][X^{\hpp},X^{\hqp}]
       G_{\hmp\hpp} G_{\hnp\hqp} \nonumber\\
&& \hspace{11em} \left.
    {}-\frac{N}{P^{(0)}_{-}} \frac{1}{2\pi l_{p}^{3}}
       \frac{i}{2} C_{m'_{1}m'_{2}m'_{3}}
       \mathcal{D}_{\tau} X^{m'_{1}}[X^{m'_{2}},X^{m'_{3}}] \right]~,
  \nonumber\\
&& \quad {} S^{(F2)}_{\mathrm{matrix}}=
   \int d\tau \, \frac{1}{N} \mathrm{Tr}
     \left[\frac{i}{2} \theta \mathcal{D}_{\tau} \theta
         +\frac{1}{2} \frac{N}{P^{(0)}_{-}}
          \frac{1}{2\pi l_{p}^{3}}
          \frac{2r}{L} \theta
          \gamma_{\hap} e^{\hap}_{\hmp}
          [X^{\hmp},\theta]  \right. \nonumber\\
 && \hspace{6em} {} +\frac{i}{2} \mathcal{D}_{\tau}X^{m'}
         \theta \left(
         \frac{1}{4} {\omega_{m'}}^{a'b'} \gamma_{a'b'}
          -\frac{1}{2L} e^{a'}_{m'} \gamma_{a'}\gamma_{\bar{r}}
          +\frac{1}{24} e^{a'}_{m'} \gamma^{b'c'd'} F_{a'b'c'd'}
          \right) \theta \nonumber\\
&& \hspace{6em} {} +\frac{1}{2} \frac{N}{P^{(0)}_{-}}
        \frac{1}{2\pi l^{3}_{p}}
        \frac{2r}{L} e^{\hap}_{\hmp} [X^{\hmp},X^{\hnp}]
        \times \nonumber\\
&& \hspace{7em} \left. {} \times  \theta \gamma_{\hap} \left(
        \frac{1}{4} {\omega_{\hnp}}^{\hbp\hcp} \gamma_{\hbp\hcp}
        +\frac{1}{4L} e^{\hbp}_{\hnp} \gamma_{\hbp\bar{r}}
        -\frac{1}{192} e^{\hbp}_{\hnp}\gamma^{c'_{1}\cdots c'_{4}}
          \gamma_{\hbp} F_{c'_{1}\cdots c'_{4}} \right)
         \theta \right]~,
   \nonumber\\
&& \quad {}S^{(F4)}_{\mathrm{matrix}}
   = -\int d\tau \, \frac{1}{N} \mathrm{Tr}
   \left[   \frac{1}{8P^{(0)}_{-}} \frac{2r}{L}
      \left[ \theta \gamma^{\hap}
             \left(\frac{1}{2L} \gamma_{\br} 
             - \frac{1}{144} \gamma^{b'_{1} \cdots b'_{4}}
               F_{b'_{1} \cdots b'_{4}} \right) \theta \right]
      \times \right. \nonumber\\
    && \hspace{17em} \times  
      \left[ \theta \gamma_{\hap} \left(
             \frac{1}{2L} \gamma_{\br}
             - \frac{1}{288} \gamma^{c'_{1} \cdots c'_{4}}
               F_{c'_{1} \cdots c'_{4}} \right) \theta \right]
     \nonumber\\
 &&  \quad  \hspace{4em}
   \left. {} + \frac{1}{384 P^{(0)}_{-}}
      \frac{2r}{L} \left( \theta \gamma^{c'_{1}c'_{2}} \theta \right)
          F_{c'_{1} \cdots c'_{4}}
          \left[ \theta \gamma^{c'_{3}c'_{4}}
                \left( \frac{1}{2L} \gamma_{\br}
                       -\frac{1}{288} \gamma^{a'_{1} \cdots a'_{4}}
                       F_{a'_{1} \cdots a'_{4}} \right) \theta
           \right] \right],
   \label{eq:matrix-membrane}
\end{eqnarray}
where the covariant derivative $\mathcal{D}_{\tau}$ means
\begin{equation}
{\left(\mathcal{D}_{\tau}X^{\hmp}\right)^{I}}_{J}
 =\partial_{\tau} {X^{\hmp I}}_{J}
   +i { [A_{\tau},X^{\hmp}]^{I}}_{J}~,
 \quad
{\left(\mathcal{D}_{\tau} \theta^{\alpha} \right)^{I}}_{J}
 =\partial_{\tau} {\theta^{\alpha\; I}}_{J}
   +i {[A_{\tau}, \theta^{\alpha}]^{I}}_{J}~.
\end{equation}
Here we have expressed the membrane tension $T$ in terms of
the eleven-dimensional Planck length $l_{p}$ by using
eq.(\ref{eq:m2tension}).
The matrix regularization (\ref{eq:matrix-reg}) entails
converting the background fields which
depend on the space-time coordinates
into the functionals of the $N \times N$ matrices
${X^{\hmp I}}_{J}$.
This gives rise to ordering ambiguity.
The problem to completely determine the ordering
is beyond the scope of this paper.
In this paper, we just adopt the ordering prescription
in ref.\cite{Myers},
which is a combination of a symmetrized trace \cite{Tseytlin},
a non-abelian Taylor expansion \cite{Garousi-Myers}
and multipole moments of currents \cite{Taylor}.

\section{D$0$-branes Propagating Near the Horizon of
D$4$-branes}
\label{sec:D0PNHD4}

In this section, we show that the action of matrix
quantum mechanics obtained in the last section
takes the same form as that of D$0$-branes propagating
near the horizon of D$4$-branes.

Coincident D$p$-branes on the curved backgrounds
are described by supersymmetric non-abelian
Born-Infeld actions coupled to background fields in curved
space-times.
Non-abelian extensions, supersymmetric extensions
and/or curved-background extensions of the Born-Infeld
actions have been carried out
\cite{Tseytlin}\cite{Bergshoeff-Townsend}\cite{CGNSW}
\cite{APS}\cite{BRS}.
We use the result of ref.\cite{Myers} for the bosonic
non-abelian D$0$-brane action on curved backgrounds.
A remarkable property of the D$p$-brane actions proposed in
ref.\cite{Myers} is that D$p$-branes can couple to the
Ramond-Ramond (R-R) potentials with the ranks higher than $p+1$,
unlike the abelian case.
{}For the fermionic sector of the D$0$-brane action,
we use the result of ref.\cite{MMS}, where
the explicit forms of the actions on bosonic curved backgrounds
are presented up to quadratic order of
the fermionic coordinate in the abelian case.

\subsection{D$4$-brane solution}

In supergravity theory, D$p$-branes are described by
black $p$-brane solutions \cite{blackbrane1}\cite{blackbrane2}.
The black $4$-brane solution 
with $Q$ unit R-R charge of type IIA supergravity
takes the following form in the string-frame:
\begin{eqnarray}
ds^{2}&=& Z(r)^{-\frac{1}{2}} \; d\hat{s}_{4+1}^{2}
          +Z(r)^{\frac{1}{2}}
             \left(dr^{2}+r^{2} d\Omega^{2}_{4}\right)~,
     \nonumber\\
e^{2\phi} &=& Z(r)^{-\frac{1}{2}}~,
  \qquad
  H_{m'_{1}\cdots m'_{4}} 
 =-\eta\; 3 \rho^{3}\epsilon^{(4)}_{m'_{1}\cdots m'_{4}}
    \sqrt{\det g(\Omega_{4})}~,
 \quad \eta = \pm 1~, \nonumber\\
 && Z(r) = 1+\frac{\rho^{3}}{r^{3}}~,
   \quad \rho = l_{s} (g_{s} Q \pi)^{\frac{1}{3}}~,
\label{eq:D4-sol}
\end{eqnarray}
where $d\hat{s}_{4+1}^{2}$ is the $(4+1)$-dimensional Minkowski
metric for the directions $(X^{0},\ldots,X^{4})$ along the D$4$-branes
and $d\Omega^{2}_{4}$ is the metric on the unit $S^{4}$:
$d\Omega^{2}_{4}=g_{m'n'}(\Omega_{4})dX^{m'}dX^{n'}$
$(m'=6,7,8,9)$.
Here $\phi$ is the dilaton and
$H_{\mu_{1}\dots\mu_{4}}$ is the field-strength
of the three-form R-R gauge field $A^{(3)}_{\mu\nu\rho}$.
The parameters
$l_{s}$ and $g_{s}$ denote the string length and
the string coupling constant, respectively.

In order to see  near-horizon geometry of the D$4$-brane
solution (\ref{eq:D4-sol}),
we take the limit
\begin{equation}
l_{s}g_{s}^{\frac{1}{3}} \rightarrow 0~,
\qquad
\mbox{with\ \ \  $\displaystyle
   \frac{r}{l_{s}^{3}g_{s}}=\mathrm{finite}$}~.
 \label{eq:limit}
\end{equation}
This limit is  essentially the same as that
taken in eq.(\ref{eq:nh-limit})
to obtain the $AdS_{7} \times S^{4}$ 
metric (\ref{eq:metric1}) from  the M$5$-brane
solution (\ref{eq:M5metric}).
This fact follows from the relation \cite{Witten}
\begin{equation}
l_{p}=l_{s}g_{s}^{\frac{1}{3}}~,
\quad
R=l_{s} g_{s}~, \label{eq:parameters}
\end{equation}
where $R$ is the radius of the direction compactified
in obtaining ten-dimensional type IIA string theory from
eleven-dimensional  M-theory.
In the limit (\ref{eq:limit}),
the D$4$-brane solution (\ref{eq:D4-sol}) becomes
\begin{eqnarray}
&\displaystyle
 ds^{2} \simeq g_{\mu\nu} dX^{\mu} dX^{\nu} 
      = \left(\frac{r}{\rho}\right)^{\frac{3}{2}}
 \; d\hat{s}_{4+1}^{2}
  + \left( \frac{\rho}{r} \right)^{\frac{3}{2}}
    \left( dr^{2}+r^{2} d\Omega_{4}^{2}\right)~,&
 \nonumber\\
&\displaystyle
e^{2\phi} \simeq \left(\frac{r}{\rho}\right)^{\frac{3}{2}}~,
\quad
H_{m'_{1}\cdots m'_{4}}
 =-\eta\; 3 \rho^{3}\epsilon^{(4)}_{m'_{1}\cdots m'_{4}}
    \sqrt{\det g(\Omega_{4})}~.&
 \label{eq:n-h-D4}
\end{eqnarray}
We note that these three variables have the structure of
$(l_{s}^{3} g_{s}) \times (\mbox{finite quantities})$.

\subsection{Bosonic sector of the
action for D$0$-branes near the horizon of D$4$-branes}

We consider the action of $N$ coincident D$0$-branes
in  the near-horizon geometry of
D$4$-branes (\ref{eq:n-h-D4}).
In this subsection,
we restrict our attention to the bosonic sector.
We will discuss the fermionic sector in the next subsection.

We make the $1+9$ split for the space-time coordinates:
$X^{\mu} = (X^{0}, X^{\hmp})$,
with $X^{\hmp}=(X^{1},\ldots,X^{4},r,X^{m'})$.
The world-line theory of $N$ coincident D$0$-branes
is a non-abelian $U(N)$ gauge theory \cite{Witten}.
We denote the gauge field by $A_{\tau}$,
where $\tau$ denotes the world-line coordinate
of the D$0$-branes.
This gauge field is accompanied by $9$ adjoint
scalar fields $\Phi^{\hmp}$ with
$\mathcal{D}_{\tau} \Phi^{\hmp}
  = \partial_{\tau} \Phi^{\hmp}
    +i [ A_{\tau}, \Phi^{\hmp}]$.
The fields $\Phi^{\hmp}$ have the dimensions of
$(\mbox{length})^{-1}$ in the space-time sense
and are related to the ($N \times N$ matrix-valued)
collective coordinates of the D$0$-branes $X^{\hmp}$ by
\begin{equation}
X^{\hmp} = 2\pi l_{s}^{2} \Phi^{\hmp}~.
\end{equation}

Substituting the background (\ref{eq:n-h-D4})
into the non-abelian D-brane action
given in ref.\cite{Myers},
we obtain the bosonic part of the action of the $N$ D$0$-branes
near the horizon of the D$4$-branes\footnote{In
ref.\cite{Ryang}, D$0$-branes in the background
(\ref{eq:n-h-D4}) is considered.
In that paper, Dielectric effects are studied and
the electric four-form field strength associated with
D$2$-brane charge is turned on.
The Chern-Simons term in the
resulting action is therefore different from
ours (\ref{eq:nonabelian-action}),
in which the magnetic four-form field strength is activated.}
in the static gauge $X^{0}=\tau$,
\begin{eqnarray}
&&S^{(B)}_{D0}\nonumber\\
&&=-T_{D0} \int d\tau \mathrm{Tr}
  \left[ e^{-\phi} \sqrt{
      -\left( g_{00} + (2\pi l_{s}^{2} )^{2}
             \mathcal{D}_{\tau} \Phi^{\hmp} g_{\hmp \hnp}
             {\left( \mathbb{Q}^{-1} \right)^{\hnp}}_{\hpp}
             \mathcal{D}_{\tau} \Phi^{\hpp}
        \right)
      \det ( \mathbb{Q}^{\hmp}_{\ \ \hnp}) }\, \right]
  \nonumber\\
&&\hspace{1em}{} - \mu_{D0} \int d\tau \mathrm{Tr}
    \left( i \frac{(2\pi l_{s}^{2})^{2}}{2}
       A^{(3)}_{m'_{1}m'_{2}m'_{3}}
       \mathcal{D}_{\tau} \Phi^{m'_{1}}
       [ \Phi^{m'_{2}} , \Phi^{m'_{3}} ]
       \right)~, 
  \label{eq:nonabelian-action}
\end{eqnarray}
where the matrix $\mathbb{Q}$ is defined as
\begin{equation}
{(\mathbb{Q})^{\hmp}}_{\hnp} = \delta^{\hmp}_{\hnp}
 + i 2\pi l_{s}^{2} [\Phi^{\hmp},\Phi^{\hpp}]
 \, g_{\hpp\hnp}~,
\end{equation}
and
$T_{D0}$ and $\mu_{D0}$ denote
the D$0$-brane tension  and charge
respectively with
\begin{equation}
T_{D0}=\mu_{D0}= \frac{1}{g_{s}l_{s}}~.
\end{equation}
In the action (\ref{eq:nonabelian-action}),
we choose the ordering prescription proposed in ref.\cite{Myers}.
Performing the $\alpha'$-expansion (with $\alpha'=l_{s}^{2}$)
in the action (\ref{eq:nonabelian-action}),
we have
\begin{eqnarray}
S^{(B)}_{D0} &=& 
 \frac{(2\pi)^{2} l_{s}^{3}}{g_{s}}
 \int d\tau
 \mathrm{Tr}
 \left[ e^{-\phi}\sqrt{-g_{00}} 
 \left( -\frac{1}{2g_{00}}
   \mathcal{D}_{\tau}\Phi^{\hmp} \mathcal{D}_{\tau} \Phi^{\hnp}
    g_{\hmp\hnp}
 \right.\right. \nonumber\\
&& \hspace{13em} \left.
     +\frac{1}{4}[\Phi^{\hmp},\Phi^{\hnp}]
     [\Phi^{\hpp},\Phi^{\hqp}]g_{\hmp\hpp}g_{\hnp\hqp}\right)
 \nonumber\\
 && \hspace{8em} \left.
   {} -\frac{i}{2} A^{(3)}_{m'_{1}m'_{2}m'_{3}}
      \mathcal{D}_{\tau} \Phi^{m'_{1}}
      [\Phi^{m'_{2}},\Phi^{m'_{3}}]
      \right]
   + \mathcal{O}(l_{s}^{5})~.
  \label{eq:alpha-leading-D0}
\end{eqnarray}
Here we have ignored the leading order contribution,
because it becomes constant in the  background (\ref{eq:n-h-D4}).

Let us compare the action (\ref{eq:alpha-leading-D0})
with the bosonic part $S^{(B)}_{\mathrm{matrix}}$
of the action (\ref{eq:matrix-membrane}).
For this purpose, we rewrite the variables of
type IIA string theory in the action (\ref{eq:alpha-leading-D0})
into those of M-theory.
Eqs.(\ref{eq:adsradius}), (\ref{eq:D4-sol})
and (\ref{eq:parameters})
yield
\begin{equation}
\rho = \frac{L}{2}~.
\end{equation}
Hence,
comparing eq.(\ref{eq:metricLC}) with eq.(\ref{eq:n-h-D4}),
we find that (see e.g.\ \cite{DHIS}\cite{Huq-Namazie}) 
\begin{eqnarray}
&&
g_{\hmp\hnp} = e^{\frac{2}{3}\phi} G_{\hmp\hnp}
  = \left(\frac{2r}{L} \right)^{\frac{1}{2}} G_{\hmp\hnp}~,
\nonumber\\
&&
H_{m'_{1}m'_{2}m'_{3}m'_{4}} = F_{m'_{1}m'_{2}m'_{3}m'_{4}}
\quad
\Leftrightarrow  \  A^{(3)}_{m'_{1}m'_{2}m'_{3}}
    =C_{m'_{1}m'_{2}m'_{3}}~.
    \label{eq:dimreduction}
\end{eqnarray}
Using these relations and eq.(\ref{eq:parameters}),
we may recast the action (\ref{eq:alpha-leading-D0}) into
\begin{eqnarray}
&&S^{(B)}_{D0}= \int d\tau \frac{1}{R}
 \mathrm{Tr}\left[
   \frac{1}{2} \frac{L}{2r}
     \mathcal{D}_{\tau} X^{\hmp} \mathcal{D}_{\tau} X^{\hnp}
     G_{\hmp \hnp}
 \right. \nonumber\\
 && \hspace{7.5em} {}
     + \frac{1}{4} \left( \frac{R}{2\pi l_{p}^{3}} \right)^{2}
       \frac{2r}{L} [X^{\hmp},X^{\hnp}][X^{\hpp},X^{\hqp}]
       G_{\hmp \hpp} G_{\hnp \hqp}
    \nonumber\\
&& \hspace{7.5em} \left.
   {} -\frac{R}{2\pi l_{p}^{3}} \frac{i}{2}
      C_{m'_{1}m'_{2}m'_{3}} \mathcal{D}_{\tau} X^{m'_{1}}
       [X^{m'_{2}},X^{m'_{3}}] \right]~.
\end{eqnarray}
This takes precisely the same form as
the action $S^{(B)}_{\mathrm{matrix}}$
with the identification
\begin{equation}
P^{(0)}_{-} = \frac{N}{R}~.
 \label{eq:P-minus0NR}
\end{equation}
Thus we have shown that
in the bosonic sector
the light-cone gauge supermembrane on $AdS_{7} \times S^{4}$
provides the matrix theory action of infinitely many D$0$-branes
propagating near the horizon of D$4$-branes.

\subsection{Fermionic Sector}
Now we turn to the fermionic sector.
In the case of a single D$0$-brane, i.e.\ the abelian
case, the explicit form  of
the fermionic sector of the Born-Infeld action
on general bosonic backgrounds
is given in ref.\cite{MMS}
up to the quadratic order of the fermionic coordinate.
We will show that this action is indeed reproduced
by the corresponding part of the
action (\ref{eq:matrix-membrane}),
i.e.\ the $U(1)$ sector of $S^{(F2)}_{\mathrm{matrix}}$,
\begin{equation}
S^{(F2)}_{U(1)}
= \int d\tau \left[\frac{i}{2}
  \theta \partial_{\tau} \theta
  + \frac{i}{2} \partial_{\tau} X^{m'}
    \theta \left( \frac{1}{4}{\omega_{m'}}^{a'b'}\gamma_{a'b'}
 -\frac{1}{2L} e^{a'}_{m'} \gamma_{a'}\gamma_{\br}
 +\frac{1}{24} e^{a'}_{m'} \gamma^{b'c'd'} F_{a'b'c'd'}
 \right)\theta \right].
 \label{eq:matrixF2U1}
\end{equation}

In order to write down the action given in ref.\cite{MMS},
we prepare notations.
On the background (\ref{eq:n-h-D4}),
the supersymmetry transformation laws
for the gravitino $\psi_{\mu}$ and the dilatino $\lambda$
of type IIA supergravity in the string-frame
take the forms\footnote{We essentially follow the  convention
in refs.\cite{typeIIsugra}\cite{MMS}.}  
\begin{equation}
\delta \psi_{\mu}
= \tilde{\nabla}_{\mu} \epsilon
=\partial_{\mu} \epsilon + W_{\mu} \epsilon~,
\quad
\delta \lambda
 = 2 \Delta \epsilon~,
\end{equation}
where
\begin{eqnarray}
W_{\mu} &=&  \frac{1}{4} \tilde{\omega}_{\mu}^{\ ab} \Gamma_{ab}
  +\frac{1}{8 \cdot 4!} e^{\phi} \tilde{e}^{b}_{\mu} 
  \Gamma^{a'_{1} \cdots a'_{4}}\Gamma_{b}
  H_{a'_{1} \cdots a'_{4}}~,
\nonumber\\
\Delta &=& \frac{1}{2}\Gamma^{a} \tilde{e}^{\mu}_{a}
         \partial_{\mu} \phi
  +\frac{1}{8 \cdot 4!} e^{\phi}
   \Gamma^{a'_{1}\cdots a'_{4}} H_{a'_{1}\cdots a'_{4}} ~,
\end{eqnarray}
with $H_{a'_{1}a'_{2}a'_{3}a'_{4}} 
      =\tilde{e}^{m'_{1}}_{a'_{1}} \tilde{e}^{m'_{2}}_{a'_{2}}
        \tilde{e}^{m'_{3}}_{a'_{3}} \tilde{e}^{m'_{4}}_{a'_{4}}
        H_{m'_{1}m'_{2}m'_{3}m'_{4}}$.
Here 
$a=(\bar{0},\hap)$ with $\hap=(\ta,\bar{r},a')$,
$\ta=(\bar{1},\ldots,\bar{4})$ and
 $a'=(\bar{6},\ldots,\bar{9})$ denote the local Lorentz indices,
 and $\tilde{\omega}_{\mu}^{\ ab}$ and $\tilde{e}^{a}_{\mu}$
are the spin-connection and the vielbein for the metric
(\ref{eq:n-h-D4}).
We introduce the pull-backs $\tilde{\nabla}_{\tau}$
and $\varrho_{\tau}$
of the covariant derivative $\tilde{\nabla}_{\mu}$
and the gamma matrices $\Gamma_{a}$
onto the D$0$-brane world-line defined as
\begin{eqnarray}
\tilde{\nabla}_{\tau} 
&=& \tilde{\nabla}_{0} + 2\pi l_{s}^{2} \partial_{\tau}
  \Phi^{\hmp} W_{\hmp}~, \nonumber\\
\varrho_{\tau}
&=& \tilde{e}^{\bar{0}}_{0}\Gamma_{\bar{0}}
  + 2\pi l_{s}^{2} \partial_{\tau}  \Phi^{\hmp} 
    \tilde{e}^{\hap}_{\hmp} \Gamma_{\hap}~.
\end{eqnarray}
Here we have chosen the static gauge $X^{0}=\tau$.
We note that
$\Phi^{\hmp}=\frac{1}{2\pi l_{s}^{2}}X^{\hmp}$
are functions commuting with each other and not matrices
because we restrict ourselves
to the $U(1)$ case.
The matrix $\varrho_{\tau}$ satisfies
\begin{equation}
(\varrho_{\tau})^{2} = \metric_{\tau\tau}~,
\end{equation}
where $\metric_{\tau\tau}$ denote the induced metric
on the D$0$-brane world-line defined as
\begin{equation}
\metric_{\tau\tau}=g_{00}
  +\left(2\pi l_{s}^{2}\right)^{2} \partial_{\tau}\Phi^{\hmp}
  \partial_{\tau} \Phi^{\hnp} g_{\hmp\hnp}~.
\end{equation}
Here we have used the fact that $g_{0\hmp}=0$ in
the background (\ref{eq:n-h-D4}).
We denote the inverse of the metric $\metric_{\tau\tau}$
by $\metric^{\tau\tau}$:
$\metric^{\tau\tau} = 1/ \metric_{\tau\tau}$.

In terms of the quantities introduced above,
the quadratic action in the fermionic sector
of the D$0$-brane is described \cite{MMS} by
\begin{equation}
S^{(F2)U(1)}_{D0}
 = \frac{i}{2} T_{D0} \int d\tau\, e^{-\phi}
   \sqrt{-\metric_{\tau\tau}}\; \bar{\Psi}
   (1-\tilde{\Gamma}_{D0}) \metric^{\tau\tau}
   \varrho_{\tau}
   \left(\tilde{\nabla}_{\tau}-\varrho_{\tau} \Delta\right)
   \Psi~,\label{eq:D0-fermion1}
\end{equation}
where $\Psi$ is a $SO(9,1)$ Majorana spinor
and $\tilde{\Gamma}_{D0}$ is defined as
\begin{equation}
\tilde{\Gamma}_{D0} = -\frac{1}{\sqrt{-\metric_{\tau\tau}}}
  \varrho_{\tau}\Gamma_{\bar{\natural}}~.
\end{equation}
Carrying out the $\alpha'$-expansion in this action,
we have
\begin{eqnarray}
S^{(F2)U(1)}_{D0}&=&
 \frac{1}{g_{s}l_{s}} \int d\tau\, e^{-\phi} \sqrt{-g_{00}}
 \left[ \frac{1}{g_{00}} i \bar{\Psi}
    \mathcal{P}^{(\mathrm{D}0)}_{+} \tilde{e}^{\bar{0}}_{0}
    \Gamma_{\bar{0}}
    \left(\partial_{\tau} + W_{0}
     - \tilde{e}^{\bar{0}}_{0} \Gamma_{\bar{0}} \Delta\right)
     \Psi \right. \nonumber\\
  && \qquad {}+ \frac{2\pi l_{s}^{2}}{g_{00}}
        \partial_{\tau} \Phi^{\hmp}
      i \bar{\Psi} \mathcal{P}_{+}^{(\mathrm{D}0)}
        \tilde{e}^{\bar{0}}_{0} \Gamma_{\bar{0}}
         \left(W_{\hmp}-\tilde{e}^{\hap}_{\hmp} \Gamma_{\hap}
         \Delta \right)\Psi \nonumber\\
  && \qquad \left. {}+\frac{2\pi l_{s}^{2}}{g_{00}}
        \partial_{\tau} \Phi^{\hmp} \frac{i}{2}
        \bar{\Psi}  \tilde{e}^{\hap}_{\hmp} \Gamma_{\hap}
          \left(\partial_{\tau}+W_{0} 
              -\tilde{e}^{\bar{0}}_{0} \Gamma_{\bar{0}}\Delta
          \right)  \Psi \right]
    + \mathcal{O}(l_{s}^{3})~,
\label{eq:D0-fermion2}
\end{eqnarray}
where $\mathcal{P}^{(\mathrm{D}0)}_{\pm}$
is the projection operator defined as
$\mathcal{P}^{(\mathrm{D}0)}_{\pm}
  = \frac{1}{2} 
    (1 \pm \Gamma_{\bar{0}}\Gamma_{\bar{\natural}} )$.

Let us compare the action (\ref{eq:D0-fermion2})
of the D$0$-brane
with the action (\ref{eq:matrixF2U1}) of the supermembrane.
For this purpose, we express the variables
of type IIA string theory in the action (\ref{eq:D0-fermion2})
by those of M-theory,
as carried out for the bosonic sector in the last subsection.
Eq.(\ref{eq:dimreduction}) enables us to choose the vielbein
$\tilde{e}^{a}_{\mu}$ such that
\begin{eqnarray}
&&\tilde{e}^{\bar{0}}_{0}
 = e^{\frac{1}{3} \phi} e^{\bar{0}}_{0}
 = \left(\frac{2r}{L}\right)^{\frac{1}{4}} e^{\bar{0}}_{0}~,
\qquad \tilde{e}^{\hap}_{\hmp}
 = e^{\frac{1}{3} \phi} e^{\hap}_{\hmp}
 =\left(\frac{2r}{L} \right)^{\frac{1}{4}}
  e^{\hap}_{\hmp}~,
\end{eqnarray}
where $e^{\ha}_{\hm}$ denotes the vielbein of  the
$AdS_{7} \times S^{4}$ metric (\ref{eq:metric1}),
the $AdS_{7}$ components of which are given
in eq.(\ref{eq:vielbein-ads7}).
This yields
\begin{eqnarray}
\frac{1}{4} \tilde{\omega}_{0}^{\ ab}\Gamma_{ab}
   &=& \frac{1}{2} \tilde{\omega}_{0}^{\ \bar{0}\bar{r}}
   = \frac{1}{2} {\omega_{0}}^{\bar{0}\bar{r}}
       \Gamma_{\bar{0}}\Gamma_{\bar{r}}
     +\frac{1}{4L} \sqrt{\frac{2r}{L}}
          \Gamma_{\bar{0}}\Gamma_{\bar{r}}~,\nonumber\\
\frac{1}{4} \tilde{\omega}_{\hmp}^{\ \; \hap\hbp}
             \Gamma_{\hap\hbp}
 &=& \frac{1}{4} {\omega_{\hmp}}^{\hap\hbp} \Gamma_{\hap\hbp}
   +\frac{1}{4L} e^{\hap}_{\hmp} \Gamma_{\hap\bar{r}}~,
  \nonumber\\
H_{a'_{1}a'_{2}a'_{3}a'_{4}}
  &=& e^{-\frac{4}{3} \phi} F_{a'_{1}a'_{2}a'_{3}a'_{4}}~,
\end{eqnarray}
where ${\omega_{\hm}}^{\ha\hb}$
denotes the spin-connection of $AdS_{7}\times S^{4}$,
the $AdS_{7}$ components of which are given
in eq.(\ref{eq:spinconnection-ads7}).
By using these relations, we find that
\begin{eqnarray}
W_{0}-\tilde{e}_{0}^{\bar{0}}\Gamma_{\bar{0}} \Delta
  &=&
W_{\tm}-\tilde{e}_{\tm}^{\ta}\Gamma_{\ta} \Delta
  = 0
  \qquad (\tm=1,\ldots,4)~,
   \nonumber\\
W_{r} - \tilde{e}^{\br}_{r} \Gamma_{\br} \Delta
  &=& {}-e^{\br}_{r} \frac{3}{4L}
  = -\frac{3}{8r}~,\nonumber\\
W_{m'}-\tilde{e}^{a'}_{m'} \Gamma_{a'} \Delta
   &=& \frac{1}{4}{\omega_{m'}}^{a'b'}\Gamma_{a'b'}
       -\frac{1}{2L} e^{a'}_{m'} \Gamma_{a'}\Gamma_{\br}
       -\frac{1}{24} e^{a'_{1}}_{m'} \Gamma^{a'_{2}a'_{3}a'_{4}}
           F_{a'_{1} \cdots a'_{4}}~.
    \label{eq:W-D}
\end{eqnarray}
By using eq.(\ref{eq:gamma-pm}), we have
\begin{equation}
\mathcal{P}^{(\mathrm{D}0)}_{\pm}
  = \mathcal{P}^{(\mathrm{LC})}_{\pm}~.
  \label{eq:proj-LCD0}
\end{equation}

Plugging eqs.(\ref{eq:W-D}) and (\ref{eq:proj-LCD0})
into the action (\ref{eq:D0-fermion2}), we obtain
\begin{eqnarray}
&& S^{(F2)U(1)}_{D0} =
 - \int d\tau\, \frac{1}{R} \left(\frac{L}{2r} \right)^{\frac{3}{4}}
   i \left[ \bar{\Psi} \mathcal{P}^{\mathrm{(LC)}}_{+} \Gamma_{\bar{0}}
            \partial_{\tau} \Psi \right. \nonumber\\
 &&\hspace{2em}
      {}+ \partial_{\tau} X^{m'} \bar{\Psi}
      \mathcal{P}^{(\mathrm{LC})}_{+} \Gamma_{\bar{0}}
      \left( \frac{1}{4}{\omega_{m'}}^{a'b'}\Gamma_{a'b'}
       -\frac{1}{2L} e^{a'}_{m'} \Gamma_{a'}\Gamma_{\br}
       -\frac{1}{24} e^{a'_{1}}_{m'} \Gamma^{a'_{2}a'_{3}a'_{4}}
           F_{a'_{1} \cdots a'_{4}}
      \right) \Psi \nonumber\\
 && \hspace{2em}  {}\left.
      +\frac{1}{2}\sqrt{\frac{L}{2r}} \partial_{\tau} X^{\hmp}
      \bar{\Psi} e^{\hap}_{\hmp} \Gamma_{\hap} \partial_{\tau} \Psi
      \right]~.
     \label{eq:D0-fermion3}
\end{eqnarray}
We note that the contribution of
$W_{r}-\tilde{e}^{\br}_{r}\Gamma_{\br} \Delta$ to this action
becomes vanishing because of eq.(\ref{eq:gamma-symmetric}).
By using the relations in Appendix \ref{sec:gamma},
we may describe the action (\ref{eq:D0-fermion3}) in terms of
the $SO(9)$ spinors $\psi^{(\uparrow)}$, $\psi^{(\downarrow)}$
and the gamma matrices $\gamma_{\hap}$ as follows:
\begin{eqnarray}
S^{(F2)U(1)}_{D0} &=& \int d\tau\, \frac{1}{R}
  \left(\frac{L}{2r}\right)^{\frac{3}{4}}
  \frac{i}{2}
   \bigg[ \psi^{(\downarrow)} \partial_{\tau}
          \psi^{(\downarrow)}
   \nonumber\\
  && \quad {}+ \partial_{\tau} X^{m'}
    \psi^{(\downarrow)} \left(
      \frac{1}{4}{\omega_{m'}}^{a'b'} \gamma_{a'b'}
      -\frac{1}{2L} e^{a'}_{m'} \gamma_{a'}\gamma_{\br}
      +\frac{1}{24} e^{a'_{1}}_{m'}
        \gamma^{a'_{2}a'_{3}a'_{4}} F_{a'_{1} \cdots a'_{4}}
        \right)
        \psi^{(\downarrow)} \nonumber\\
   && \quad -\frac{1}{2} \sqrt{\frac{L}{2r}}
      \partial_{\tau} X^{\hmp} e_{\hmp}^{\hap}
      \left( \psi^{(\downarrow)} \gamma_{\hap}
             \partial_{\tau} \psi^{(\uparrow)}
             +\psi^{(\uparrow)} \gamma_{\hap}
                \partial_{\tau} \psi^{(\downarrow)} \right)
      \bigg]~.
\end{eqnarray}
Here we have decomposed the Majorana spinor $\Psi$ as
\begin{equation}
\Psi = \frac{1}{\sqrt{2}}
       \left( \begin{array}{c}
               \psi^{(\uparrow)} \\ \psi^{(\downarrow)}
              \end{array}
       \right)~.
\end{equation}
The supersymmetric D$p$-brane action possesses
the fermionic gauge symmetry ($\kappa$-symmetry).
In the present case, under the $\kappa$-transformation,
the fermionic coordinate $\Psi$ transforms \cite{MMS} as
\begin{equation}
\delta_{\kappa} \Psi
 =(1-\tilde{\Gamma}_{D0}) \kappa + \mathcal{O}(\Psi^{3})
 =(1+\Gamma_{\bar{0}}\Gamma_{\bar{\natural}}) \kappa
  + \mathcal{O}(l_{s}^{2}, \Psi^{3})~,
\end{equation}
where the transformation parameter $\kappa$ is
a $\tau$-dependent $SO(9,1)$ Majorana spinor.
We may therefore impose the following
condition\footnote{From eq.(\ref{eq:proj-LCD0}),
    we find that the gauge condition (\ref{eq:D0-gauge})
    formally takes the same form as the light-cone gauge condition
    (\ref{eq:LCfermi}) imposed on the fermionic coordinate
    $\Theta$ of the supermembrane.}
on the fermionic coordinate $\Psi$
to fix this gauge symmetry,
\begin{equation}
\mathcal{P}^{(\mathrm{D}0)}_{+} \Psi=0
\quad \Leftrightarrow \quad
\psi^{(\uparrow)}=0~.
\label{eq:D0-gauge}
\end{equation}
In this gauge, the last term in the action (\ref{eq:D0-fermion3})
becomes vanishing.
We thus find that the action (\ref{eq:D0-fermion3})
takes the same form as the action (\ref{eq:matrixF2U1})
with the identification
\begin{equation}
\psi^{(\downarrow)}
 =\sqrt{\frac{R}{N}} \left(\frac{2r}{L}\right)^{\frac{3}{8}}
 \theta~,
 \label{eq:psidownarrow}
\end{equation}
with $N=1$.
As addressed in Section \ref{sec:redefinition},
the normalization factor of the fermionic
coordinate $\theta$ may be absorbed into conventions,
even though it depends on the bosonic fields.
In this way, we have verified that,
as well as the bosonic sector, the fermionic sector
of the matrix action of the light-cone supermembrane on
$AdS_{7}\times S^{4}$
has the interpretation as the Matrix theory action
of D$0$-branes.

\section{Conclusions and Discussions}
 \label{sec:summary}

In this paper, we obtained matrix quantum mechanics
for a supermembrane on $AdS_{7} \times S^{4}$ from
the light-cone supermembrane on this background.
We constructed light-cone gauge
formulation for the supermembrane on $AdS_{7} \times S^{4}$
in a similar way to the flat case \cite{dWHN}.
Taking account of the fact that $AdS_{7} \times S^{4}$
is obtained as near-horizon geometry of the M$5$-brane solution
\cite{Gibbons-Townsend}\cite{Maldacena},
we expect that the resulting matrix quantum mechanics
(\ref{eq:matrix-membrane}) should govern the system
of D$0$-branes propagating near the horizon
of the D$4$-brane solution of type IIA supergravity.
We verified that the action (\ref{eq:matrix-membrane})
has such a Matrix theory interpretation.
We showed that the bosonic part $S^{(B)}_{\mathrm{matrix}}$
of the matrix action (\ref{eq:matrix-membrane})
takes the same form as the leading order terms in
the $\alpha'$-expansion of the non-abelian Born-Infeld action
\cite{Myers} of D$0$-branes.
As for the fermionic sector, we compared the action
(\ref{eq:matrix-membrane}) with the results in ref.\cite{MMS},
where the explicit form of the action is presented
up to quadratic order
in the fermionic coordinate $\theta$ for a single D$p$-brane
on a general bosonic curved background.
We showed that the corresponding part of our action
(\ref{eq:matrixF2U1}) reproduces their results.

The fermionic sector of the action (\ref{eq:matrix-membrane})
also contains non-abelian parts
and quartic order terms of $\theta$.
{}From the fact that the action (\ref{eq:matrix-membrane})
coincides with the part of the D$0$-brane action stated
in the above, we may expect that our result should give a hint
as to the non-abelian extension of
the fermionic sectors of the Born-Infeld actions
for D$0$-branes on curved backgrounds.
We may rewrite the action $S^{(F2)}_{\mathrm{matrix}}$
in eq.(\ref{eq:matrix-membrane}) in terms of variables
of ten-dimensional type IIA string theory into the following
form in the gauge (\ref{eq:D0-gauge}):
\begin{eqnarray}
\lefteqn{
S^{(F2)}_{\mathrm{matrix}} =
  \frac{1}{g_{s}l_{s}} \int d\tau \mathrm{Tr}
  \Bigg[ e^{-\phi} \sqrt{-g_{00}} \times
} \nonumber\\
 && \quad \times i\left(\frac{1}{g_{00}} \bar{\Psi}
    \mathcal{P}^{(\mathrm{D}0)}_{+} \tilde{e}^{\bar{0}}_{0}
    \Gamma_{\bar{0}}
    \left(\mathcal{D}_{\tau}+W_{0}
          -\tilde{e}^{\bar{0}}_{0} \Gamma_{\bar{0}} \Delta \right)
    \Psi -\bar{\Psi} \mathcal{P}^{(\mathrm{D}0)}_{+}
    \Gamma_{\bar{\natural}} \Gamma{\hap} \tilde{e}^{\hap}_{\hmp}
    i [\Phi^{\hmp},\Psi] \right. \nonumber\\
  && \quad \qquad {}+\frac{2\pi l_{s}^{2}}{g_{00}}
     \mathcal{D}_{\tau} \Phi^{\hmp} \bar{\Psi}
     \mathcal{P}^{(\mathrm{D}0)}_{+}
     \tilde{e}^{\bar{0}}_{0}\Gamma_{\bar{0}}
       \left( W_{\hmp} - \tilde{e}^{\hap}_{\hmp} \Gamma_{\hap}
              \Delta\right) \Psi    \nonumber\\
  && \quad \qquad {}
      +\frac{2\pi l_{s}^{2}}{g_{00}}
      \mathcal{D}_{\tau} \Phi^{\hmp} \frac{1}{2}
      \bar{\Psi} \tilde{e}^{\hap}_{\hmp} \Gamma_{\hap}
      \left( \mathcal{D}_{\tau} + W_{0} 
              - \tilde{e}^{\bar{0}}_{0} \Gamma_{\bar{0}} \Delta
      \right) \Psi     \nonumber\\
   && \left. \quad \qquad {}- 2\pi l_{s}^{2}
        \tilde{e}^{\hap}_{\hmp} i [\Phi^{\hmp},\Phi^{\hnp}]
        \bar{\Psi} \mathcal{P}^{(\mathrm{D}0)}_{+}
        \Gamma_{\bar{\natural}} \Gamma_{\hap} W_{\hnp} \Psi
       \right) \Bigg]~,
       \label{eq:covariantize}
\end{eqnarray}
with the identifications (\ref{eq:P-minus0NR})
and (\ref{eq:psidownarrow}).
We note that there are ambiguities in covariantizing
the $SO(9)$ spinor $\theta$ and
gamma matrices $\gamma_{\hap}$
into the $SO(9,1)$ spinor $\Psi$
and gamma matrices $\Gamma_{\ha}$ in eq.(\ref{eq:covariantize}).
The first, the third and the fourth terms on the r.h.s.\ in
eq.(\ref{eq:covariantize}) are directly obtained from
the $U(1)$ part of the D$0$-brane action (\ref{eq:D0-fermion2})
by replacing  $\partial_{\tau} \Phi^{\hmp}$
with $\mathcal{D}_{\tau} \Phi^{\hmp}$.
This replacement is a part of steps in the non-abelian extension
of the Born-Infeld actions \cite{Myers}.
We may naturally guess that the last term on the r.h.s.\ in
eq.(\ref{eq:covariantize}) should be derived from
terms proportional to the field strength $F_{\mu\nu}$
of the gauge field $A_{\mu}$ in the D$p$-brane action
via T-duality.
We make a comment on the four-Fermi terms $S^{(F4)}_{\mathrm{matrix}}$
of the action (\ref{eq:matrix-membrane}).
In Section \ref{sec:LCgauge},
we used the Fierz relations (\ref{eq:id-1}) and (\ref{eq:id-2})
in the four-Fermi terms, before applying the matrix regularization
to the membrane action.
Whether we are allowed to use such Fierz relations also after
the matrix regularization
depends on the ordering prescription
in the fermionic sector of the non-abelian Born-Infeld action.
This point needs further investigations.

In this paper, we did not discuss the (super-)isometry
of $AdS_{7}\times S^{4}$.
It is because the isometries of the target spaces
of light-cone supermembranes
become subtle after the matrix regularization.
In fact, even in the flat case, the eleven-dimensional
Lorentz symmetry is obscure in the matrix action
\cite{dWMN}\cite{EMM}\cite{Melosch}.
Such obscurity originates in the fact that
it is not known how to employ the matrix regularization
for the coordinate $X^{-}$ and consequently for the Lorentz
generators $M^{-\hap}$, since $X^{-}$ explicitly depends
on the geometry of $\Sigma_{(2)}$ in the light-cone gauge.

As mentioned in Introduction, supermembrane theory on
$AdS_{7} \times S^{4}$ is supposed to be dual to
six-dimensional superconformal field theory
\cite{Witten95}\cite{Strominger}\cite{Witten5-brane}\cite{Seiberg}.
We hope, therefore, that our matrix action should provide a
useful new approach to the study of the $AdS_{7}/CFT_{6}$
correspondence beyond the supergravity level and
shed light on this problem.

\subsection*{Acknowledgements}
The author is grateful to E.~Sezgin for discussions
on supermembranes on $AdS \times S$.
This work has been supported in part by NSF Grant
PHY-0314712.

\vspace{5ex}

\appendix

\section{$SO(10,1)$ and $SO(9)$ gamma matrices}
\label{sec:gamma}

The $SO(10,1)$ gamma matrices $\Gamma_{\ha}$
$(\ha=\bar{0},\ldots,\bar{9},\bar{\natural})$
satisfy the $SO(10,1)$ Clifford algebra
\begin{eqnarray}
\Gamma_{\ha}\Gamma_{\hb} + \Gamma_{\hb}\Gamma_{\ha}
 = 2 \eta_{\ha\hb}~,
\quad
\eta_{\ha\hb} =  \mathrm{diag} (-1,1,\ldots, 1)~.
\end{eqnarray}
The hermitian conjugation of $\Gamma_{\ha}$ is given by
$\Gamma^{\dagger}_{\ha}
 = \Gamma_{\bar{0}} \Gamma_{\ha} \Gamma_{\bar{0}}$.
The charge conjugation of the $SO(10,1)$ spinor
$\Psi$ is defined as
$\Psi^{c} = \mathcal{C} \bar{\Psi}^{T}$,
where $\bar{\Psi}=\Psi^{\dagger} \Gamma_{\bar{0}}$
is the Dirac conjugate of $\Psi$
and $\mathcal{C}$ is the charge conjugation matrix defined by
\begin{equation}
\Gamma^{T}_{\ha}=-\mathcal{C}^{-1} \Gamma_{\ha} \mathcal{C}~,
\quad
\mathcal{C}^{T}=-\mathcal{C}~.
\label{eq:c-matrix}
\end{equation}
The Majorana spinor $\Psi_{\mathrm{M}}$
is defined by $\Psi_{\mathrm{M}}^{c}=\Psi_{\mathrm{M}}$.
This leads to
\begin{equation}
\bar{\Psi}_{\mathrm{M}}
 = - \Psi_{\mathrm{M}}^{T} \mathcal{C}^{-1}~.
\end{equation}

Eq.(\ref{eq:c-matrix}) yields
\begin{equation}
\left( \mathcal{C}^{-1} \Gamma_{\ha_{1}\cdots \ha_{n}}\right)^{T}
=(-1)^{\frac{1}{2}(n+2)(n-1)}\,
  \mathcal{C}^{-1} \Gamma_{\ha_{1}\cdots \ha_{n}}~.
\end{equation}
This implies that the matrix
$\mathcal{C}^{-1} \Gamma_{\ha_{1}\cdots \ha_{n}}$
is symmetric for $n\equiv 1,2$ (mod $4$)
and anti-symmetric for $n\equiv 0,3$ (mod $4$).
Thereby we obtain
\begin{equation}
\bar{\Psi}_{\mathrm{M}}\Gamma_{\ha_{1} \cdots \ha_{n}}
\Psi_{\mathrm{M}} =0
\quad \mbox{for $n \equiv 1,2 \pmod{4}$}~,
\label{eq:gamma-symmetric}
\end{equation}
where $\Psi_{\mathrm{M}}$ is an arbitrary Majorana spinor.

A set of the matrices
\begin{equation}
(\mathcal{C}^{-1})_{\hat{\alpha}\hat{\beta}}~,
\quad
(\mathcal{C}^{-1}\Gamma_{\ha})_{\hat{\alpha}\hat{\beta}}~,
\quad
(\mathcal{C}^{-1}\Gamma_{\ha_{1}\ha_{2}})_{\hat{\alpha}\hat{\beta}}~,
\ \cdots \ ,\ (\mathcal{C}^{-1}\Gamma_{\ha_{1}\cdots \ha_{5}}
  )_{\hat{\alpha}\hat{\beta}}
\end{equation}
composes a complete basis of the $32\times 32$
matrices.
The completeness relation reads
\begin{equation}
\delta^{\hat{\epsilon}}_{\hat{\alpha}}
\delta^{\hat{\delta}}_{\hat{\beta}}
  = \frac{1}{32}\sum_{p=0}^{5} \frac{1}{p!}
    (\Gamma^{\ha_{p}\cdots \ha_{1}}\mathcal{C}
    )^{\hat{\delta}\hat{\epsilon}}
    (\mathcal{C}^{-1}\Gamma_{\ha_{1}\cdots \ha_{p}}
    )_{\hat{\alpha}\hat{\beta}}~.
\end{equation}
Using this relation, we can prove the identities
\begin{eqnarray}
 && (\mathcal{C}^{-1}\Gamma_{\ha\hb}
       )_{(\hat{\alpha}\hat{\beta}}
   \,
   (\mathcal{C}^{-1}\Gamma^{\hb}
        )_{\hat{\delta}\hat{\epsilon})}
    =0~, \label{eq:Fierz1} \\
 &&
(\mathcal{C}^{-1} \Gamma_{\ha\hb}
     )_{\hat{\alpha}[\hat{\beta}}
   \, (\mathcal{C}^{-1}\Gamma^{\ha\hb\hc}
       )_{\hat{\delta}\hat{\epsilon}]}
   -6 (\mathcal{C}^{-1}\Gamma^{\hc}
        )_{\hat{\alpha}[\hat{\beta}}
     \, (\mathcal{C}^{-1})_{\hat{\delta}\hat{\epsilon}]}
     =0~.
  \label{eq:Fierz2}
\end{eqnarray}

In this paper, we choose the Majorana representation
for the $SO(10,1)$ spinors
in which the Majorana spinors become real spinors.
One can find that the following relations hold
in this representation:
\begin{equation}
\mathcal{C}=\Gamma_{\bar{0}}~,
\quad
\Gamma_{\ha}^{\ast}=\Gamma_{\ha}~,
\label{eq:cgamma1}
\end{equation}
where $\Gamma^{\ast}_{\ha}$ denotes the complex conjugate of
$\Gamma_{\ha}$.

In accordance with the light-cone coordinates (\ref{eq:LC}),
we introduce $(\Gamma^{\bpu},\Gamma^{\bmi})$ defined as
\begin{equation}
\Gamma^{\bpu} = \Gamma_{\bmi}
=\frac{1}{\sqrt{2}}
  \left(\Gamma^{\bar{\natural}} + \Gamma^{\bar{0}}\right)~,
\quad
\Gamma^{\bmi}=\Gamma_{\bpu}
=\frac{1}{\sqrt{2}}
  \left( \Gamma^{\bar{\natural}} -\Gamma^{\bar{0}} \right)~.
\end{equation}
These matrices obey the relations
\begin{equation}
\Gamma^{\bpu}\Gamma^{\bmi} + \Gamma^{\bmi} \Gamma^{\bpu}
 = 2\eta^{\bpu\bmi} =2~,
\quad (\Gamma^{\bpu})^{2}=\eta^{\bpu\bpu}=0~,
\quad (\Gamma^{\bmi})^{2}=\eta^{\bmi\bmi}=0~.
\end{equation}

We decompose the gamma matrices $\Gamma_{\ha}$ into
$SO(1,1) \times SO(9)$ gamma matrices as follows:
\begin{eqnarray}
&&\Gamma_{\bar{0}}=-i\sigma_{2} \otimes \mathbf{1}_{16}
        =\left( \begin{array}{cc}
                 \displaystyle
                 0 & -\mathbf{1}_{16} \\ \mathbf{1}_{16} & 0
                \end{array} \right)~,
\quad \Gamma_{\bar{\natural}} = -\sigma_{1} \otimes \mathbf{1}_{16}
        = \left( \begin{array}{cc}
                   0 & -\mathbf{1}_{16} \\
                   -\mathbf{1}_{16} & 0
                  \end{array}
          \right)~,\nonumber\\
&& {} \Gamma_{\hap}=\sigma_{3} \otimes \gamma_{\hap}
    = \left( \begin{array}{cc}
              \gamma_{\hap} & 0 \\
               0 & -\gamma_{\hap}
              \end{array} \right)
     \qquad \hap =( \bar{1}, \ldots, \bar{4},\bar{r},
                    \bar{6},\ldots,\bar{9})~,
 \label{eq:decompose}
\end{eqnarray}
where $(\sigma_{1},\sigma_{2},\sigma_{3})$ denote the standard
Pauli matrices and $\gamma_{\hap}$ denote $SO(9)$ gamma
matrices, which satisfy the $SO(9)$ Clifford algebra,
\begin{equation}
\gamma_{\hap} \gamma_{\hbp} + \gamma_{\hbp} \gamma_{\hap}
 = 2 \delta_{\hap\hbp}~.
\end{equation}
Using eqs.(\ref{eq:c-matrix}) and (\ref{eq:cgamma1}),
we find that
the $SO(9)$ gamma matrices $\gamma_{\hap}$ in eq.(\ref{eq:decompose})
are real and symmetric:
\begin{equation}
\gamma_{\hap}^{T} = \gamma_{\hap}~, \quad
\gamma_{\hap}^{\ast}= \gamma_{\hap}~.
\end{equation}
The $SO(9)$ charge conjugation
matrix $\mathcal{C}_{9}$, defined by
$\gamma^{T}_{\hap}
= \mathcal{C}_{9}^{-1} \gamma_{\hap} \mathcal{C}_{9}$
and $\mathcal{C}_{9}^{T}=\mathcal{C}_{9}$,
may hence be chosen to be unity, $\mathcal{C}_{9}=1$, in this
representation.

In accordance with
the decomposition (\ref{eq:decompose}) of the
gamma matrices, an arbitrary $SO(9,1)$ spinor
$\Psi$ is decomposed into two $SO(9)$ spinors as
\begin{equation}
\Psi^{\hat{\alpha}} = \left(
  \begin{array}{c}
    \psi^{(\uparrow)\alpha} \\ \psi^{(\downarrow)\alpha}
   \end{array} \right)
   \qquad (\alpha=1,\ldots,16)~.
   \label{eq:ue-shita}
\end{equation}
Eq.(\ref{eq:decompose}) yields
\begin{eqnarray}
&&\Gamma^{\bpu} = \left(
   \begin{array}{cc} 0&0\\ -\sqrt{2}&0 \end{array}
  \right)~, \quad
 \Gamma^{\bmi} = \left(
   \begin{array}{cc} 0& -\sqrt{2} \\ 0&0 \end{array}
   \right)~,
  \nonumber\\
  && \mathcal{P}^{\mathrm{(LC)}}_{-}
    =\frac{1-\Gamma^{\bmi\bpu}}{2}
    =\left(
    \begin{array}{cl} 0&0\\ 0& \mathbf{1}_{16} \end{array}
     \right)~,
    \quad
      \mathcal{P}_{+}^{(\mathrm{LC})}
        =\frac{1+\Gamma^{\bmi\bpu}}{2}
        =\left(
    \begin{array}{lc} \mathbf{1}_{16}&0\\ 0&0  \end{array}
     \right)~,
\end{eqnarray}
where
\begin{equation}
   \Gamma^{\bmi\bpu}
    = \frac{1}{2}
       (\Gamma^{\bmi}\Gamma^{\bpu} - \Gamma^{\bpu}\Gamma^{\bmi})
    =\Gamma_{\bar{0}}\Gamma_{\bar{\natural}}~.
   \label{eq:gamma-pm}
\end{equation}
This leads to
\begin{equation}
\Psi_{-} 
\equiv \mathcal{P}^{(\mathrm{LC})}_{-} \Psi
=\left( \begin{array}{c}
         0 \\ \psi^{(\downarrow)}
        \end{array}  \right)~,
\quad
\Psi_{+}
\equiv  \mathcal{P}_{+}^{(\mathrm{+})} \Psi
= \left( \begin{array}{cc}
           \psi^{(\uparrow)} \\ 0
          \end{array} \right)~.
 \label{eq:halfspinor}
\end{equation}

\section{Derivation of eq.(\ref{eq:4-fermi})}
 \label{sec:derivation}

In this appendix, we derive eq.(\ref{eq:4-fermi}).
In the following, we will frequently use the relations
(\ref{eq:gamma-theta}), (\ref{eq:P-gamma})
and (\ref{eq:gamma-symmetric}),
without mentioning it.

We begin by providing relations for later use.
Multiplying eq.(\ref{eq:Fierz2}) by
$\Theta_{-}^{\hat{\beta}} \Theta_{-}^{\hat{\delta}}
 \Theta_{-}^{\hat{\epsilon}}$ 
and setting $\hc=\bmi$,
we have
\begin{equation}
\left( \bar{\Theta}_{-} \Gamma^{\bmi}\Gamma^{\hap\hbp} \Theta_{-}
\right) \bar{\Theta}_{-} \Gamma_{\hap\hbp} =0~.
\label{eq:id-1}
\end{equation}
Multiplying eq.(\ref{eq:Fierz1}) by
$(\Gamma^{\bmi}\Theta_{-})^{\hat{\alpha}}
 \, \Theta_{-}^{\hat{\beta}}
 \, (\Gamma_{c'_{2}c'_{3}c'_{4}} \Theta_{-})^{\hat{\delta}}
 F^{c'_{1}c'_{2}c'_{3}c'_{4}}$
and setting $\ha=c'_{1}$, we may show that
\begin{eqnarray}
\lefteqn{
\left( \bar{\Theta}_{-} \Gamma^{\bmi} \Gamma_{c'_{1}\hbp}
       \Theta_{-} \right)
   \bar{\Theta}_{-} \Gamma_{c'_{2}c'_{3}c'_{4}}
     \Gamma^{\hbp} F^{c'_{1} \cdots c'_{4}}
   }
   \nonumber\\
  &&={}-\left( \bar{\Theta}_{-} \Gamma^{\bmi}
           \Gamma_{c'_{2}c'_{3}c'_{4}} \Theta_{-} \right)
     \bar{\Theta}_{-} \Gamma_{c'_{1}} F^{c'_{1} \cdots c'_{4}}
   + 3 \left( \bar{\Theta}_{-} \Gamma^{\bmi}
         \Gamma_{c'_{3}c'_{4}} \Theta_{-} \right)
      \bar{\Theta}_{-} \Gamma_{c'_{1}c'_{2}}
            F^{c'_{1} \cdots c'_{4}}~.
     \label{eq:id-2}
\end{eqnarray}
Combining the identities
\begin{eqnarray}
\Gamma^{\hap\hbp} \Gamma_{c'_{1}c'_{2}c'_{3}c'_{4}}
 &=& {\Gamma^{\hap\hbp}}_{c'_{1} \cdots c'_{2}}
 -8\delta^{[\hap}_{[c'_{1}}  {\Gamma^{\hbp]}}_{c'_{2}c'_{3}c'_{4}]}
 -12 \delta^{\hap}_{[c'_{1}} \delta^{\hbp}_{c'_{2}}
     \Gamma_{c'_{3}c'_{4}]}~, \nonumber\\
\Gamma_{c'_{2}c'_{3}c'_{4}} \Gamma^{\hbp}
  &=& -{\Gamma^{\hbp}}_{c'_{2}c'_{3}c'_{4}}
      +3 \delta^{\hbp}_{[c'_{2}} \Gamma_{c'_{3}c'_{4}]}~,
\end{eqnarray}
we obtain
\begin{equation}
{\Gamma^{\hap\hbp}}_{c'_{1}c'_{2}c'_{3}c'_{4}}
 = \Gamma^{\hap\hbp} \Gamma_{c'_{1}c'_{2}c'_{3}c'_{4}}
   -8 \delta^{[\hap}_{[c'_{1}}
      \Gamma_{c'_{2}c'_{3}c'_{4}]}  \Gamma^{\hbp]}
   + 36 \delta^{\hap}_{[c'_{1}} \delta^{\hbp}_{c'_{2}}
        \Gamma_{c'_{3}c'_{4}]}~.
    \label{eq:simple}
\end{equation}

Eq.(\ref{eq:M2-LC}) yields
\begin{eqnarray}
\bar{\Theta}_{-} \Gamma^{\bmi}
  \mathcal{M}^{2} [\Theta_{-}]
  &=& \frac{i}{18} \left(\bar{\Theta}_{-}
      \Gamma_{b'_{2}b'_{3}b'_{4}} F^{b'_{1} \cdots b'_{4}}
      \Theta_{-} \right)
      \bar{\Theta}_{-} \Gamma_{b'_{1}} 
  \nonumber\\
  && {} + \frac{i}{288} \left( \bar{\Theta}_{-}
           \Gamma^{\bmi} \Gamma_{\hap\hbp} \Theta_{-} \right)
         \bar{\Theta}_{-}
         \Gamma^{\hap\hbp c'_{1}\cdots c'_{4}}
         F_{c'_{1} \cdots c'_{4}} \nonumber\\
   && {}+ \frac{i}{12} \left( \bar{\Theta}_{-} \Gamma^{\bmi}
          \Gamma_{c'_{1}c'_{2}} \Theta_{-} \right)
          \bar{\Theta}_{-} \Gamma_{c'_{3}c'_{4}}
           F^{c'_{1} \cdots c'_{4}}~.
     \label{eq:sono1}
\end{eqnarray}
Here we have used the following relation,
which follows from eq.(\ref{eq:gamma-symmetric}):
\begin{equation}
\bar{\Theta}_{-} \Gamma^{\bmi}
 {\Gamma_{\hap}}^{b'_{1} \cdots b'_{4}} \Theta_{-} =0~.
 \label{eq:kantan}
\end{equation}
Substituting eq.(\ref{eq:simple}) into the second term
on the r.h.s.\ in eq.(\ref{eq:sono1}), we have
\begin{eqnarray}
&&
\bar{\Theta}_{-} \Gamma^{\bmi}
\mathcal{M}^{2} [\Theta_{-}]
\label{eq:sono2} \\
  &&= \frac{i}{18} \left(\bar{\Theta}_{-}
      \Gamma_{b'_{2}b'_{3}b'_{4}} F^{b'_{1} \cdots b'_{4}}
      \Theta_{-} \right)
      \bar{\Theta}_{-} \Gamma_{b'_{1}} 
  + \frac{i}{288} \left( \bar{\Theta}_{-}
           \Gamma^{\bmi} \Gamma_{\hap\hbp} \Theta_{-} \right)
         \bar{\Theta}_{-}
         \Gamma^{\hap\hbp} \Gamma_{c'_{1}\cdots c'_{4}}
         F^{c'_{1} \cdots c'_{4}} \nonumber\\
  &&\quad {} -\frac{i}{36} \left( \bar{\Theta}_{-} \Gamma^{\bmi}
         \Gamma_{c'_{1} \hbp} \Theta_{-} \right)
        \bar{\Theta}_{-} \Gamma_{c'_{2}c'_{3}c'_{4}}
        \Gamma^{\hbp} F^{c'_{1} \cdots c'_{4}}
    + i \frac{5}{24} \left( \bar{\Theta}_{-} \Gamma^{\bmi}
          \Gamma_{c'_{1}c'_{2}} \Theta_{-} \right)
          \bar{\Theta}_{-} \Gamma_{c'_{3}c'_{4}}
           F^{c'_{1} \cdots c'_{4}}~. \nonumber
\end{eqnarray}
The second term on the r.h.s.\ in this equation
is vanishing because of eq.(\ref{eq:id-1}).
Plugging eq.(\ref{eq:id-2}) into the third term
on the r.h.s.\ in eq.(\ref{eq:sono2}),
we obtain
\begin{equation}
\bar{\Theta}_{-} \Gamma^{\bmi}
\mathcal{M}^{2} [\Theta_{-}] 
=\frac{i}{12} \left(\bar{\Theta}_{-}
      \Gamma_{b'_{2}b'_{3}b'_{4}} F^{b'_{1} \cdots b'_{4}}
      \Theta_{-} \right)
      \bar{\Theta}_{-} \Gamma_{b'_{1}}
 + \frac{i}{8} \left( \bar{\Theta}_{-} \Gamma^{\bmi}
          \Gamma_{c'_{1}c'_{2}} \Theta_{-} \right)
          \bar{\Theta}_{-} \Gamma_{c'_{3}c'_{4}}
           F^{c'_{1} \cdots c'_{4}}~.
    \label{eq:sono3}
\end{equation}
By using eq.(\ref{eq:kantan}), we may recast this equation
into
\begin{eqnarray}
\bar{\Theta}_{-} \Gamma^{\bmi}
\mathcal{M}^{2} [\Theta_{-}] 
&=& \frac{i}{48} \left( \bar{\Theta}_{-} \Gamma^{\bmi}
    \Gamma^{\hap} \Gamma_{b'_{1} \cdots b'_{4}}
     F^{b'_{1} \cdots b'_{4}} \Theta_{-} \right)
     \bar{\Theta}_{-} \Gamma_{\hap} 
   \nonumber\\
&& {}+ \frac{i}{8} \left( \bar{\Theta}_{-} \Gamma^{\bmi}
          \Gamma_{c'_{1}c'_{2}} \Theta_{-} \right)
          \bar{\Theta}_{-} \Gamma_{c'_{3}c'_{4}}
           F^{c'_{1} \cdots c'_{4}}~.
    \label{eq:sono4}
\end{eqnarray}
Substituting this equation into the
definition  (\ref{eq:def-4-fermi}) of
$\mathbb{F}$,
we obtain eq.(\ref{eq:4-fermi}).
Thus we have derived eq.(\ref{eq:4-fermi}).

\section{Explicit Forms of Several Quantities}
\label{sec:note}

In this appendix, we substitute the explicit form of
the field $F_{a'b'c'd'}$ in the $AdS_{7}\times S^{4}$
solution (\ref{eq:ads7-sol}) into
 several quantities in the text.
As mentioned in Section \ref{sec:LCgauge},
it is convenient to use the fermionic coordinates
$(\lamm,\psm)$ introduced in eq.(\ref{eq:m5-projected}).
We note that the `chirality' associated with the projection
operator $\mathcal{P}^{(\mathrm{M}5)}_{\pm}$
flips under the Dirac conjugation:
\begin{equation}
\bar{\Theta}^{(\pm)}_{-}
  = \bar{\Theta}^{(\pm)}_{-}
     \mathcal{P}^{(\mathrm{M}5)}_{\mp}~.
\end{equation}

Plugging eq.(\ref{eq:ads7-sol}) into eq.(\ref{eq:M2-LC}),
we find that $\mathcal{M}^{2}[\Theta_{-}]$ takes the form
\begin{eqnarray}
&&\hspace*{-1em}
  {\left(
  \mathcal{M}^{2}[\Theta_{-}]\right)^{\hat{\alpha}}}_{\hat{\beta}}
= \eta \frac{i}{L} \left[
     \left(\Gamma_{\ta}\hga \Theta_{-}\right)^{\hat{\alpha}}
      \left(\bar{\Theta}_{-} \Gamma^{\ta}\right)_{\hat{\beta}}
      +\left(\Gamma_{\br} \hga \Theta_{-}\right)^{\hat{\alpha}}
    \left(\bar{\Theta}_{-} \Gamma^{\br}\right)_{\hat{\beta}}
    -2 \left( \Gamma_{a'}\hga \Theta_{-}
                       \right)^{\hat{\alpha}}
       \left(\bar{\Theta}_{-} \Gamma^{a'} \right)_{\hat{\beta}}
  \right. \nonumber \\
&&\hspace{7.5em}
    {}- \Theta_{-}^{\hat{\alpha}}
       \left(\bar{\Theta}_{-} \hga \right)_{\hat{\beta}}
    - \left( \Gamma_{\br} \Gamma_{\ta} \Theta_{-}\right)^{\hat{\alpha}}
      \left(\bar{\Theta}_{-}\Gamma^{\br}\Gamma^{\ta} \hga
        \right)_{\hat{\beta}}
      \nonumber\\
&&\hspace{7.5em} \left.
    {}-\frac{1}{2}\left(\Gamma_{\ta \tb} \Theta_{-} 
                  \right)^{\hat{\alpha}}
           \left(\bar{\Theta}_{-} \Gamma^{\ta\tb} \hga
            \right)_{\hat{\beta}}
     + \left(\Gamma_{a'b'}\Theta_{-}\right)^{\hat{\alpha}}
       \left(\bar{\Theta}_{-} \Gamma^{a'b'} \hga \right)_{\hat{\beta}}
     \right]~,
\end{eqnarray}
where $\hga$ is the matrix defined in eq.(\ref{eq:M5-projector}).
{}From eq.(\ref{eq:d-minus}), we obtain
\begin{eqnarray}
\tilde{D}_{i} \Theta_{-} 
&=& \partial_{i} \left(\lamm + \psm \right)
    \nonumber\\
&&{} + \partial_{i}X^{+} e^{\bpu}_{+} \frac{1}{L}
        \Gamma^{\bmi} \Gamma_{\br} \psm 
     + \partial_{i} X^{\tm} e^{\ta}_{\tm} \frac{1}{L}
        \Gamma_{\ta}\Gamma_{\br} \psm
 + \partial_{i}r \frac{1}{4r} \left(\psm-\lamm\right)
 \nonumber\\
&&{}+\partial_{i} X^{m'}
   \left(\frac{1}{4} {\omega_{m'}}^{a'b'}\Gamma_{a'b'}
         +\frac{\eta}{L} e^{a'}_{m'} \Gamma_{a'} \hga \right)
   \left(\lamm+\psm\right)~.
   \label{eq:m2dLC}
\end{eqnarray}
Using these relations, we obtain a detailed form of the
pull-back $\Pi^{A}_{i}$ of the supervielbein.
The component $\Pi^{\bmi}_{i}$ is expressed as
\begin{eqnarray}
\lefteqn{\Pi^{\bmi}_{i} =
   \partial_{i} X^{-} e^{\bmi}_{-}
   -i \blamm \Gamma^{\bmi} \partial_{i} \lamm
   -i \bpsm \Gamma^{\bmi} \partial_{i} \psm
 } \nonumber\\
 &&{} -i\partial_{i}X^{\tm} e^{\ta}_{\tm} \frac{1}{L}
      \blamm \Gamma^{\bmi} 
      \Gamma_{\ta} \Gamma_{\br} \psm
   \nonumber\\
  && {}-i\partial_{i} X^{m'} \left[
     \blamm \Gamma^{\bmi} \left( \frac{1}{4} {\omega_{m'}}^{a'b'}
      \Gamma_{a'b'}
      + \frac{\eta}{L} e^{a'}_{m'} \Gamma_{a'} \hga \right) \lamm
      \right. \nonumber\\
  && \hspace{5em} \left. {}+ \bpsm
      \Gamma^{\bmi} \left( \frac{1}{4} {\omega_{m'}}^{a'b'}
      \Gamma_{a'b'}
      + \frac{\eta}{L} e^{a'}_{m'} \Gamma_{a'} \hga \right)
      \psm \right] \nonumber\\
  && {}- \partial_{i} X^{+} e^{\bpu}_{+} \frac{1}{8L^{2}}
        \left[ 2 \left( \bpsm \Gamma^{\bmi} \Gamma_{a'}
                   \Gamma_{\br} \psm
                   - \blamm \Gamma^{\bmi} \Gamma_{a'} \Gamma_{\br}
                      \lamm \right)
                  \bpsm \Gamma^{\bmi} \Gamma^{a'} 
                        \Gamma_{\br} \psm  \right.
       \nonumber\\
   && \hspace{6em} \quad \left. {}+\left(
           \bpsm \Gamma^{\bmi} \Gamma_{a'b'} \psm
            + \blamm \Gamma^{\bmi} \Gamma_{a'b'} \lamm
              \right)
              \bpsm \Gamma^{\bmi} \Gamma^{a'b'} \psm
       \right]~.
\end{eqnarray}
Here we have used eq.(\ref{eq:sono3}).
The components
$\Pi^{\ta}_{\tau}$ and $\Pi^{a'}_{\tau}$ take the forms
\begin{eqnarray}
\Pi^{\ta}_{\tau} &=& \partial_{i}X^{\tm} e^{\ta}_{\tm}
  -i  e^{\bpu}_{+}\frac{1}{L}
    \blamm \Gamma^{\ta}\Gamma^{\bmi}\Gamma_{\br} \psm~,
 \nonumber\\
\Pi^{a'}_{\tau} &=& \partial_{\tau}X^{m'} e^{a'}_{m'}
  -i e^{\bpu}_{+}  \frac{1}{L}
       \bpsm \Gamma^{a'} \Gamma^{\bmi} \Gamma_{\br} \psm~.
\end{eqnarray}
The fermionic components $\Pi_{\tau}$ and $\Pi_{\acute{\imath}}$
become
\begin{eqnarray}
 \Pi_{\tau} &=& \partial_{\tau} \left( \lamm + \psm \right)
 \nonumber\\
 &&{}+  e^{\bpu}_{+} \frac{1}{L}
           \Gamma^{\bmi} \Gamma_{\br} \psm
         + \partial_{\tau} X^{\tm} e^{\ta}_{\tm}
           \Gamma_{\ta} \Gamma_{\br} \psm
      - \partial_{i}r e^{\br}_{r}\frac{1}{2L}
         \left(\psm -\lamm\right)
     \nonumber\\
 &&{} + \partial_{i}X^{m'}
         \left( \frac{1}{4} {\omega_{m'}}^{a'b'} \Gamma_{a'b'}
                +\frac{\eta}{L} e^{a'}_{m'} \Gamma_{a'} \hga
         \right) \left( \lamm + \psm \right)
        \nonumber\\
&&{}+ e^{\bpu}_{+} \frac{1}{6L^{2}} 
          \left[
             2\Gamma_{\br}\Gamma_{\ta} \psm
             \left( \blamm \Gamma^{\bmi}\Gamma^{\br}\Gamma^{\ta}
                    \psm \right) \right.
         \nonumber\\
    && \hspace{5em}
        {}-2 \Gamma_{\br}\Gamma_{a'}\left(\psm-\lamm \right)
            \left(\bpsm \Gamma^{\bmi}\Gamma^{\br}\Gamma^{a'}
                  \psm\right)
         \nonumber\\
    && \hspace{5em}
          {}+\frac{1}{2} \Gamma_{\ta\tb} \left(\lamm +\psm \right)
             \left(\bpsm \Gamma^{\bmi}\Gamma^{\ta\tb} \psm \right)
        \nonumber\\
   && \hspace{5em} \left.
          {} -\Gamma_{a'b'}\left(\lamm + \psm \right)
             \left( \bpsm \Gamma^{\bmi}\Gamma^{a'b'} \psm\right)
           \right]~,\nonumber\\
  \Pi_{\acute{\imath}} 
    &=& \partial_{\acute{\imath}} 
        \left( \lamm + \psm \right)\nonumber\\
    &&{}  + \partial_{\acute{\imath}}
          X^{\tm} e^{\ta}_{\tm}
           \Gamma_{\ta} \Gamma_{\br} \psm
       - \partial_{\acute{\imath}}r  e^{\br}_{r}
            \frac{1}{2L} \left(\psm -\lamm\right)
     \nonumber\\
   &&{}
     + \partial_{i}X^{m'}
         \left( \frac{1}{4} {\omega_{m'}}^{a'b'} \Gamma_{a'b'}
                +\frac{\eta}{L} e^{a'}_{m'} \Gamma_{a'} \hga
         \right) \left( \lamm + \psm \right)~.
\end{eqnarray}

The elements $h_{\tau\tau}$ and
$u_{\acute{\imath}}\equiv h_{\tau \acute{\imath}}$
of the induced metric $h_{ij}$ turn out to be
\begin{eqnarray}
h_{\tau\tau} &=& 2\partial_{\tau} X^{-} G_{+-}
   \partial_{\tau}X^{\hmp} \partial_{\tau} X^{\hnp}
    G_{\hmp \hnp}  \nonumber\\
  &&{}-i 2 e^{\bpu}_{+}
     \left( \blamm \Gamma^{\bmi}\partial_{\tau} \lamm
           + \bpsm \Gamma^{\bmi} \partial_{\tau} \psm \right)
   \nonumber\\
  && {} -i2 \partial_{\tau} X^{m'} e^{\bpu}_{+}
     \left[ \blamm \Gamma^{\bmi}
     \left(\frac{1}{4} {\omega_{m'}}^{a'b'} \Gamma_{a'b'}
           + \frac{\eta}{L}e^{a'}_{m'} \Gamma_{a'}\hga
     \right) \lamm \right. \nonumber\\
  && \hspace{7em} \left. {}+ \bpsm \Gamma^{\bmi}
     \left(\frac{1}{4} {\omega_{m'}}^{a'b'} \Gamma_{a'b'}
           + 2 \frac{\eta}{L}e^{a'}_{m'} \Gamma_{a'}\hga
     \right) \psm \right]
 \nonumber\\
  && {}+ \mathbb{F}~, \nonumber\\
u_{\acute{\imath}}
&=&  \partial_{\acute{\imath}} X^{-} G_{+-}
      +\partial_{\tau} X^{\hmp} \partial_{\tau} X^{\hnp}
       G_{\hmp\hnp} \nonumber\\
&& {}-i e^{\bpu}_{+}
      \left(\blamm \Gamma^{\bmi} \partial_{\acute{\imath}} \lamm
            + \bpsm \Gamma^{\bmi} \partial_{\acute{\imath}} \psm \right)
    \nonumber\\
&& {} -i  e^{\bpu}_{+}
      \partial_{\acute{\imath}} X^{m'}
      \left[ \blamm \Gamma^{\bmi}
         \left( \frac{1}{4} {\omega_{m'}}^{a'b'} \Gamma_{a'b'}
                 +\frac{\eta}{L} e^{a'}_{m'}  \Gamma_{a'} \hga\right)
         \lamm \right. \nonumber \\
   && \hspace{8em} \left.{}+ \bpsm \Gamma^{\bmi}
          \left( \frac{1}{4} {\omega_{m'}}^{a'b'} \Gamma_{a'b'}
                 +2\frac{\eta}{L} e^{a'}_{m'}  \Gamma_{a'} \hga\right)
           \psm \right]~,
\end{eqnarray}
where $\mathbb{F}$ is the four-Fermi term introduced in
eq.(\ref{eq:def-4-fermi}).
{}From eq.(\ref{eq:4-fermi}), we find that
this is described by
\begin{eqnarray}
\mathbb{F}
&=& -\frac{(e^{\bpu}_{+})^{2}}{L^{2}}
  \left[ \left(\blamm \Gamma^{\bmi} \Gamma_{\ta}\Gamma_{\br}
               \psm\right)
         \left( \blamm \Gamma^{\bmi}\Gamma^{\ta}\Gamma_{\br}
              \psm \right) \right.
   \nonumber\\
&& \hspace{3em} {}+\frac{1}{2} \left(
    3 \bpsm \Gamma^{\bmi} \Gamma_{a'} \Gamma_{\br} \psm
    - \blamm \Gamma^{\bmi}\Gamma_{a'} \Gamma_{\br} \lamm
    \right)
    \left( \bpsm \Gamma^{\bmi} \Gamma^{a'} \Gamma_{\br}
           \psm \right)
    \nonumber\\
&& \hspace{3em} \left.  {}+\frac{1}{4} \left(
    \bpsm \Gamma^{\bmi} \Gamma_{a'b'} \psm
    + \blamm \Gamma^{\bmi} \Gamma_{a'b'} \lamm
    \right)
    \left( \bpsm \Gamma^{\bmi} \Gamma^{a'b'} \psm \right)
    \right]~.
\end{eqnarray}

Let us denote
the $\Theta$-dependent part of the Wess-Zumino term
by $\mathcal{L}^{(F)}_{\mathrm{WZ}}$,
\begin{eqnarray}
\mathcal{L}^{(F)}_{\mathrm{WZ}}
&\equiv &{} -T i \int^{1}_{0} dt \epsilon^{ijk} \bar{\Theta}_{-}
    \Gamma_{\ha \hb} \Pi_{i} (X,t\Theta_{-})
    \Pi^{\ha}_{j}(X,t\Theta_{-})
    \Pi^{\hb}_{k}(X,t\Theta_{-}) \nonumber\\
&=& T i \epsilon^{\acute{\imath}\acute{\jmath}}
    e^{\bpu}_{+} e^{\hap}_{\hmp}
     \partial_{\acute{\imath}} X^{\hmp}
     \bar{\Theta}_{-} \Gamma^{\bmi} \Gamma_{\hap}
       \left[ \partial_{\acute{\jmath}} \Theta_{-}
   + \partial_{\acute{\jmath}}X^{\hnp}
    \left( \frac{1}{4} {\omega_{\hnp}}^{\hbp \hcp}
           \Gamma_{\hbp \hcp}
      \right. \right. \nonumber\\
   &&\hspace{7em}\left. \left.
       {}+\frac{1}{4L} e^{\hbp}_{\hnp} \Gamma_{\hbp\br}
            +\frac{1}{192} e^{\hbp}_{\hnp}
              \Gamma^{c'_{1} \cdots c'_{4}} \Gamma_{\hbp}
              F_{c'_{1} \cdots c'_{4}} \right) \Theta_{-}
        \right]~.
\end{eqnarray}
This is expressed as
\begin{eqnarray}
\mathcal{L}^{(F)}_{\mathrm{WZ}}
&=&T i  \epsilon^{\acute{\imath}\acute{\jmath}}
      \left[ \partial_{\acute{\imath}}X^{\tm}
             e^{\bpu}_{+} e^{\ta}_{\tm}
       \left( \blamm \Gamma^{\bmi}\Gamma_{\ta}
       \partial_{\acute{\jmath}}\psm
       +\bpsm \Gamma^{\bmi}\Gamma_{ta}
       \partial_{\acute{\jmath}}\lamm\right)
      \right. \nonumber\\
   && \hspace{2em} {}+ \partial_{\acute{\imath}}r
       \, e^{\bpu}_{+} e^{\br}_{r}
      \left(\blamm \Gamma^{\bmi}\Gamma_{\br} 
               \partial_{\acute{\jmath}}\lamm
            +\bpsm \Gamma^{\bmi}\Gamma_{\br}
               \partial_{\acute{\jmath}}\psm \right)
      \nonumber\\
   && \hspace{2em} {}+\partial_{\acute{\imath}} X^{m'}
      e^{\bpu}_{+} e^{a'}_{m'}
      \left(\blamm \Gamma^{\bmi}\Gamma_{a'}
             \partial_{\acute{\jmath}} \lamm
         +\bpsm \Gamma^{\bmi}\Gamma_{a'}
             \partial_{\acute{\jmath}} \psm \right)
         \nonumber\\
   && \hspace{2em} {}+ \partial_{\acute{\imath}}X^{\tm}
                       \partial_{\acute{\jmath}}X^{\tn}
          \frac{3}{2L} e^{\bpu}_{+} e^{\ta}_{\tm} e^{\tb}_{\tn}
       \, \bpsm \Gamma^{\bmi} \Gamma_{\ta \tb} \Gamma_{\br}
          \psm 
       \nonumber\\
    && \hspace{2em}
       {}+\partial_{\acute{\imath}}X^{\tm}\partial_{\acute{\jmath}}r
        \frac{3}{L} e^{\bpu}_{+} e^{\ta}_{\tm}e^{\br}_{r}
       \blamm \Gamma^{\bmi} \Gamma_{\ta} \psm
      \nonumber\\
    && \hspace{2em} {}+ 
      \partial_{\acute{\imath}}X^{\tm}
      \partial_{\acute{\jmath}}X^{n'}
      2e^{\bpu}_{+} e^{\ta}_{\tm}
       \bpsm \Gamma^{\bmi} \Gamma_{\ta}
       \left(\frac{1}{4} {\omega_{n'}}^{a'b'} \Gamma_{a'b'}
             +\frac{\eta}{L} e^{a'}_{n'} \Gamma_{a'} \hga\right)
           \lamm \nonumber\\
    && \hspace{2em}
      {}- \partial_{\acute{\imath}}X^{m'}
          \partial_{\acute{\jmath}}r\,
           e^{\bpu}_{+} e^{\br}_{r}\,
          \frac{1}{4} {\omega_{m'}}^{a'b'}
        \left(\blamm \Gamma^{\bmi}\Gamma_{\br}\Gamma_{a'b'} \lamm
         +\bpsm \Gamma^{\bmi} \Gamma_{\br} \Gamma_{a'b'} \psm \right)
       \nonumber\\
    && \hspace{2em} {}+ 
       \partial_{\acute{\imath}}X^{m'} \partial_{\acute{\jmath}}X^{n'}
         e^{\bpu}_{+} e^{a'}_{m'}
         \left\{ \blamm \Gamma^{\bmi} \Gamma_{a'}
          \left(\frac{1}{4}{\omega_{n'}}^{c'd'} \Gamma_{c'd'}
                +\frac{\eta}{L} e^{c'}_{n'} \Gamma_{c'} \hga \right)
          \lamm \right.
          \nonumber\\
        && \hspace{8em}\left.\left.
          {}+\bpsm \Gamma^{\bmi} \Gamma_{a'}
          \left( \frac{1}{4}{\omega_{n'}}^{c'd'} \Gamma_{c'd'}
             +\frac{\eta}{2L} e^{c'}_{n'} \Gamma_{c'} \hga \right)
            \psm \right\}\right].
\end{eqnarray}
Here we have used the relation
${\omega_{\tm}}^{\ta\br}=\frac{1}{L} e^{\ta}_{\tm}$.
Using the notation $\mathcal{L}^{(F)}_{\mathrm{WZ}}$,
we may write the hamiltonian (\ref{eq:hamiltonian0})
as
\begin{equation}
\mathcal{H}_{0}
= \frac{G_{+-}}{2P_{-}}
    \left[ T^{2} \bar{h}
        +P_{\tm}G^{\tm\tn}P_{\tn}
        +P_{r}G^{rr}P_{r} 
        + Q_{m'}G^{m'n'} Q_{n'}
        \right]
     -\mathcal{L}^{(F)}_{\mathrm{WZ}}
 -\frac{P_{-}}{2G_{+-}} \mathbb{F}~,
\end{equation}
where $Q_{m'}$ is given in eq.(\ref{eq:q}),
which is described as
\begin{eqnarray}
Q_{m'}
&=&P_{m'}-\frac{T}{2} \epsilon^{\acute{\imath}\acute{\jmath}}
              \partial_{\acute{\imath}}X^{m'_{1}}
              \partial_{\acute{\jmath}}X^{m'_{2}}
              C_{m'm'_{1}m'_{2}} 
       \nonumber\\
  && {}+i\frac{P_{-}}{e^{\bmi}_{-}}
      \left[
         \blamm \Gamma^{\bmi}\left(
          \frac{1}{4} {\omega_{m'}}^{a'b'}\Gamma_{a'b'}
           +\frac{\eta}{L}e^{a'}_{m'}\Gamma_{a'} \hga \right)
           \lamm
        \right.\nonumber\\
   && \hspace{4em} \left. {}
           +\bpsm \Gamma^{\bmi} \left(
            \frac{1}{4} {\omega_{m'}}^{a'b'}\Gamma_{a'b'}
            +2\frac{\eta}{L}e^{a'}_{m'}\Gamma_{a'} \hga \right)
               \psm \right]~.
\end{eqnarray}


\end{document}